\documentclass[12pt]{article}

\usepackage[default]{jasa_harvard}    
\usepackage{JASA_manu}

\usepackage{amsmath}
\usepackage{graphicx,psfrag,epsf}
\usepackage{subfigure,amsmath,latexsym,amssymb}
\usepackage{float,epsfig,multirow,rotating,times}
\usepackage{upgreek,wrapfig}
\usepackage{algorithm}
\usepackage{enumerate}

\newcommand {\ctn}{\citeasnoun} 

\newcommand{\topline}{\hrule height 1pt width \textwidth \vspace*{2pt}}
\newcommand{\botline}{\vspace*{2pt}\hrule height 1pt width \textwidth \vspace*{4pt}}
\newtheorem{algo}{Algorithm} 
\numberwithin{algo}{section}

\newcommand{\bY}{\boldsymbol{Y}}

\begin{document}




\normalsize

\title{\vspace{-0.8in}{Perfect Simulation for Mixtures with Known and Unknown Number of Components}}
\author{Sabyasachi Mukhopadhyay and 
Sourabh Bhattacharya\footnote{Sabyasachi Mukhopadhyay is a PhD student and Sourabh Bhattacharya is an
Assistant Professor in Bayesian and Interdisciplinary 
Research Unit, Indian Statistical Institute, Kolkata, India. Corresponding e-mail: {\it sourabh@isical.ac.in}}}
\date{\vspace{-0.5in}}
\maketitle

\begin{abstract}
We propose and develop a novel and effective perfect sampling methodology for simulating
from posteriors corresponding to mixtures with either known (fixed) or unknown number of components.
For the latter we consider the Dirichlet process-based mixture model developed by these authors,
and show that our ideas are applicable to conjugate, and importantly, to non-conjugate cases.
As to be expected, and as we show, perfect sampling for mixtures with known number of components
can be achieved with much less effort with a simplified version of our general methodology, whether or not
conjugate or non-conjugate priors are used. While no special assumption
is necessary in the conjugate set-up for our theory to work, we require the assumption of compact
parameter space in the non-conjugate set-up.
However, we argue, with appropriate
analytical, simulation, and real data studies as support, that such compactness assumption is not
unrealistic and is not an impediment in practice.
Not only do we validate our ideas theoretically and with simulation studies, but we also consider
application of our proposal to three real data sets used by several authors in the past
in connection with mixture models. The results we achieved in each of our experiments
with either simulation study or real data application, are quite encouraging. However,
the computation can be extremely burdensome in the case of large number of mixture components
and in massive data sets. We discuss the role of parallel processing in mitigating the
extreme computational burden.

\noindent%
{\it Keywords:}
Bounding chains; Dirichlet process; Gibbs sampling; Mixtures; Optimization; Perfect Sampling

\end{abstract}

\section{Introduction}
\label{sec:introduction}

Markov chain Monte Carlo (MCMC) algorithms are developed to simulate
from desired distributions, from which generation of exact samples is difficult.
The methodology has found much use in the Bayesian statistical
paradigm thanks to the natural need to sample from intractable posterior distributions.
But in whatever clever way the
MCMC algorithms are designed, the samples are generated only approximately.
Due to impossibility of running the chain for an infinite span of time, a suitable burn-in period is chosen,
usually by a combination of empirical and ad-hoc means.
The realizations retained after discarding the burn-in period are presumed to
closely represent the true distribution. The degree of closeness, however,
depends upon how suitably the burn-in is chosen, and an arbitrary choice may lead to serious bias.
Even in simple problems non-negligible biases often result if the burn-in period is chosen
inadequately (see, for example, \ctn{Roberts98}). Such problems can only be aggravated in the case of realistic, more complex models, such
as mixture models of the form, given for the data point $y$, by
\begin{equation}
[y\mid\Theta_p,\Pi_p]=\sum_{j=1}^p\pi_jf(y\mid\theta_j),
\label{eq:mixture}
\end{equation}
In (\ref{eq:mixture}), $\Theta_p$ denotes the set of parameters $(\theta_1,\ldots,\theta_p)'$,  $\Pi_p=(\pi_1,\ldots,\pi_p)'$ are the mixing probabilities such that $\pi_j>0$ for
$j=1,\ldots,p$, and $\sum_{j=1}^p\pi_j=1$. Here the number of mixture components $p$ may or may not be known.
The latter case corresponds to variable dimensional parameter space since the cardinality of the set
$\Theta_p$ then becomes random.

Mixture models form a very important class of models in statistics, known for their
versatility. The Bayesian paradigm even allows for random number of mixture components
(making the dimensionality of the parameter space a random variable),
adding to the flexibility of mixture models. Sophisticated MCMC algorithms are needed
for posterior inference in mixture models, raising the question of adequacy of the available practical
convergence assessment methods, particularly in the case of
variable-dimensional mixture models. The importance of the aforementioned class of models
makes it important to solve the associated convergence assessment problem.
In this paper, we develop a rigorous solution to this problem using the principle of perfect sampling.

The perfect sampling methodology, first proposed in the seminal paper by \ctn{Prop96}, attempts to
completely avoid the problems of MCMC convergence assessment.
In principle, starting at all possible initial values, so many parallel Markov chains need to be run, each
starting at time $t= -\infty$. If by time $t=0$, all the chains coalesce, the coalescent point at time $t=0$ is an exact realization
from the stationary distribution. Essentially, this principle works in the same way as the regular MCMC algorithms,
but by replacing its starting time $t=0$ with $t= -\infty$ and the convergence time $t=\infty$ with $t=0$.
To achieve perfect sampling in practice, \ctn{Prop96} proposed the ``coupling from the past" (CFTP) algorithm,
which avoids running Markov chains from the infinite past.
We briefly describe this in the next section.

\section{The CFTP algorithm}

For the time being, for the sake of clarity, following \ctn{Prop96}, let us assume that the
state space $\mathcal X$ is finite,
and let $\{X_t;t=0,1,\ldots\}$ denote the underlying Markov chain. Then, for $t\geq 0$
it is possible to represent the Markov chain generically as a random mapping:
$X_{t+1}=\phi_t(X_t)=\phi(X_t,R_{t+1})$, for some
function $\phi(\cdot,\cdot)$ and an $iid$ sequence $\{R_t;t=1,\ldots\}$.
Then the CFTP algorithm is as follows (see \ctn{Prop96}, \ctn{Robert04}):
\begin{itemize}
\item[1.] For $t=-1,-2,\ldots$, generate $\phi_t(x)$ for $x\in\mathcal X$.
\item[2.] For $t=-1,-2,\ldots$, for $x\in\mathcal X$, define the compositions
\begin{equation}
\Phi_t(x)=\phi_0\circ\phi_{-1}\circ\cdots\phi_{-t}(x)
\label{eq:composition}
\end{equation}
\item[3.] Determine the time $T$ such that $\Phi_T$ is constant.
\item[4.] Accept $\Phi_T(x^*)$ as an exact realization from the
stationary distribution for any arbitrary $x^*\in\mathcal X$.
\end{itemize}
It is well-known (see, for example, \ctn{Casella01})
that the above algorithm terminates almost surely in finite time under
very mild conditions and
indeed yields a realization distributed exactly
according to the stationary distribution of the Markov chain.
\ctn{Prop96} recommend taking $t=-2^j$, for $j=1,2,\ldots$, which we shall
adopt in this paper. A subtle, but important point is that, even if all the Markov chains
coalesce before time $t=0$, the corresponding simulation at the time of coalescence need not yield
a perfect sample. One needs to carry the algorithm forward till time $t=0$; the sample
corresponding to only $t=0$ is guaranteed to be perfect. For details, see \ctn{Casella01}.

Although in the seminal paper of \ctn{Prop96} the CFTP algorithm, as described above,
was constructed assuming finite state space in the above algorithm, later developments
managed to circumvent this assumption of finiteness.
Indeed, strategies for perfect sampling in general state
spaces are described in \ctn{Murdoch98} and \ctn{Green99}, but quite restricted set-ups,
which do not hold generally, are needed
to implement such strategies. The set up of mixture models is much complex, and the known strategies
are difficult to apply.

The first attempt to construct perfect sampling algorithms for mixture
models is by \ctn{Hobert99}. However, they assumed only 2-component and 3-component mixture
models, where only the mixing probabilities are assumed to be unknown. Bounding chains
(this term seems to first appear in \ctn{Huber98}; \ctn{Huber04} also uses this term) with monotonicity
structures are used to enable the CFTP algorithm in these cases. Using the principle of the perfect slice sampler
(\ctn{Mira01}), and assuming conjugate priors on the parameters, \ctn{Casella02} proposed
a perfect sampling methodology for mixtures with known number
of components by marginalizing out the parameters.
It is noted in \ctn{Casella02} that in the conjugate case the marginalized form of the
posterior is analytically available, but the authors point out (see Section 2 of \ctn{Casella02})
that still perfect simulation from the analytically available marginalized posterior is important.
Unfortunately, apart from the somewhat restricted assumptions
of conjugate priors and known number of components, the methodology is approximate in nature
and the authors themselves demonstrated that the approximation can be quite poor.
\ctn{Fearnhead05} proposed a direct sampling methodology based on recursion relations
associated with the forward-backward algorithm, for mixtures of discrete
distributions assuming a conjugate set-up and known number of components, thus bringing in
an extra and crucial assumption of discrete data. Most recently, \ctn{Breyer10} introduced
a new perfect sampling methodology in mixtures with known number
of components where only the mixture weights are unknown. The method is also shown to work
for normal mixtures (with known number of components) with unknown weights as well as with
unknown means, but with a known, common variance. However, even in this much restricted
setup, the practicality of the algorithm is challenged. Indeed, the authors honestly remarked
in pages 255--256 the following: ``Unfortunately, in practice this extension of our
algorithm seems to be feasible only for fairly small data sets $(n < 5)$ where the data consist
of one cluster."

However, the drawbacks
of the methodologies in no way present the contributions of the aforementioned authors
in poor light, these only show how difficult the problem is.
In this paper we attempt to avoid the restrictions and difficulties by proposing a novel approach.
In the non-conjugate case (but not in the conjugate case) we are forced to assume compactness
of the parameter space,
but we argue in Section \ref{subsec:bound_known_k}, followed up with
a simulated data example in the supplement and three real data cases in Section \ref{sec:real_application},
that it is not an unrealistic
assumption, particularly in the Bayesian paradigm. Noting particularly that no methodology
exists in the literature that even attempts perfect simulation from mixtures with unknown number
of components, for either compact or non-compact parameter space, for either conjugate
or non-conjugate set-up, there is no reason to look upon
our compactness assumption only in the non-conjugate case as a serious drawback.

We first construct a perfect sampling algorithm for mixture models with fixed (known)
number of components and then generalize the ideas to mixtures with unknown number of components.
For the sake of illustration, we concentrate on mixtures of normal densities, but our ideas are
quite generally applicable.
We illustrate our methodology with simulation studies as well as with application to three real
data sets. Additional technical details and further
details on experiments are provided in the supplement, whose
sections and figures have the prefix ``S-'' when referred to in this paper.

\section{Perfect sampling for normal mixtures with known number of components}
\label{sec:normixture_known_k}

\subsection{Normal mixture model and prior distributions}
\label{subsec:model_prior}

Letting $f(\cdot\mid\theta_j)$ in (\ref{eq:mixture}) denote normal densities with mean $\mu_j$
and variance $\sigma^2_j$, we obtain the following
normal mixture model
\begin{equation}
[y\mid\Theta_p,\Pi_p]=\sum_{j=1}^p\pi_j\sqrt{\frac{\lambda_j}{2\pi}}\exp\left\{-\frac{\lambda_j}{2}\left(y-\mu_j\right)^2\right\},
\label{eq:normixture}
\end{equation}
In (\ref{eq:normixture}), $\theta_j=(\mu_j,\lambda_j)$, where $\lambda_j=\sigma^{-2}_j$.
For the sake of convenience of illustration only we consider the following conjugate prior specification on the unknown variables
\begin{eqnarray}
\lambda_j&\stackrel{iid}{\sim}&Gamma(\eta/2, \zeta/2);j=1,\ldots,p\label{eq:prior_lambda}\\
\left[\mu_j\mid \lambda_j\right]&\stackrel{iid}{\sim}& N(\xi_j, \tau_j^2 \lambda^{-1}_j); j=1,\ldots,p\label{eq:prior_mu}\\
\Pi_p=(\pi_1,\ldots,\pi_p)&\sim& Dirichlet(\gamma_1, \ldots, \gamma_p)\label{eq:prior_pi}\\
\end{eqnarray}
In (\ref{eq:prior_lambda}), $Gamma(\eta/2,\zeta/2)$ denotes the $Gamma$ distribution with density
proportional to \\$\lambda^{{\eta/2}-1}_j\exp\left\{-\zeta\lambda_j/2\right\}$.
We further assume that $\{\eta,\zeta\}$, $\{\xi_1,\ldots,\xi_p\}$, $\{\tau_1,\ldots,\tau_p\}$ and $\{\gamma_1,\ldots,\gamma_p\}$
are known.

With conjugate priors the marginal posteriors of the parameters $(\Pi_p,\Theta_p)$ and
the allocation variables $Z$ are available in closed forms, but still sampling from the posterior distributions
is important. Indeed, \ctn{Casella02} argue that sampling enables inference on arbitrary functionals
of the unknown variables, which are not analytically available.
These authors proposed a perfect slice sampler for sampling from the marginal posterior of the
allocation variable $Z$
only. Given perfect samples from the posterior of $Z$, drawing exact samples from the posterior distributions of
$(\Pi_p,\Theta_p)$ is straightforward.
But importantly, the posteriors are not available in closed forms in non-conjugate situations,
and even Gibbs sampling
is not straightforward in such cases. Since our goal is
to provide a general theory that works for both conjugate and non-conjugate priors, we do not focus on the
marginalized approach, although the conjugate situation is just a special (and simpler) case of our
proposed principle
(see Sections \ref{subsec:bound_known_k} and \ref{subsec:sb_bounding_chains}).
Due to convenience of illustration we begin with the conjugate prior case
where the full conditional distributions needed for Gibbs sampling are available.
It will be shown how the same ideas are carried over to the non-conjugate cases.

\subsection{Full conditional distributions}
\label{subsec:fullcond}

Assuming that a dataset $Y=(y_1,\ldots,y_n)'$ is available, let us define the set of allocation variables $Z=(z_1,\ldots,z_n)'$, where
$z_i= j$ if $y_i$ comes from the $j$-th component of the mixture. Further, defining $n_j=\#\{i:z_i=j\}$, $\bar y_j = \sum_{i:z_i=j}y_i/n_j$,
$Z_{-i}=(z_1,\ldots,z_{i-1},z_{i+1},\ldots,z_n)'$ and $\Theta_{-jp}=(\theta_1,\ldots,\theta_{j-1},\theta_{j+1},\ldots,\theta_p)'$,
the full conditional distributions of the unknown random variables can be expressed as the following:
{\small
\begin{eqnarray}
[z_i=j\mid \Theta_p, Z_{-i}, \Pi, Y]&\propto& \pi_j\sqrt{\lambda_j}\exp\left\{-\frac{\lambda_j}{2}(y_i-\mu_j)^2\right\}
\label{eq:full_cond_z}\\
\left[\lambda_j\mid Z, \Pi, \Theta_{-jp},\mu_j, Y\right] &\sim& Gamma\left(\frac{\eta+n_j}{2},\frac{1}{2}
\left\{\zeta+\frac{n_j(\bar y_j-\xi_j)^2}{n_j\tau^2_j+1}+\sum_{i:z_i=j}(y_i-\bar y_j)^2\right\}\right)\nonumber\\
\label{eq:full_cond_lambda}\\
\left[\mu_j\mid \Theta_{-jp},\lambda_j,Z,\Pi,Y\right]&\sim&N\left(\frac{n_j\bar y_j\tau_j^2+\xi_j}{n_j\tau_j^2+1}, \frac{\tau^2_j}{\lambda_j\left(n_j\tau_j^{2}+1\right)}\right) \label{eq:full_cond_mu}\\
\left[\Pi\mid Z, \Theta, Y\right]&\sim& Dirichlet\left(n_1+\gamma_1,\ldots,n_p+\gamma_p\right) \label{eq:full_cond_pi}
\end{eqnarray}
}

Perfect sampling, making use of the full conditional distributions available for Gibbs sampling, has been developed by \ctn{Moller99}. But
the development is based on the assumption that the random variables are discrete and that the distribution functions are monotonic in the conditioned variables.
These are not satisfied in the case of mixtures. Full conditional based perfect sampling has also been used by \ctn{Schneider04} in the context of Bayesian
variable selection in a linear regression model, but their methods depend strongly on the underlying structure of their
linear regression model and prior assumptions and do not apply to mixture models.
Our proposed method hinges on obtaining stochastic lower and upper bounds for the $Z$-part of the Gibbs sampler, and simulating only from
the two bounding chains, and noting their coalescence. It turns out that, in our methodology, there is no need to simulate the other unknowns,
$(\Pi_p,\Theta_p)$
before coalescence, even in the non-conjugate set-up.
Details are provided in the next section.

\subsection{Bounding chains for $Z$}
\label{subsec:bound_known_k}

For $i=1,\ldots,n$, let $F_i(\cdot\mid Y,Z_{-i},\Pi_p,\Theta_p)$ denote the distribution function corresponding to the full conditional of $z_i$.
Writing $X_{-i}=(Z_{-i},\Pi_p,\Theta_p)$, let
\begin{eqnarray}
F^L_i(\cdot\mid Y)&=&\inf_{X_{-i}}F_i(\cdot\mid Y,X_{-i})\label{eq:lower}\\
F^U_i(\cdot\mid Y)&=&\sup_{X_{-i}}F_i(\cdot\mid Y,X_{-i})\label{eq:upper}
\end{eqnarray}
be the lower and the upper bounds of  $F_i(\cdot\mid Y,Z_{-i},\Pi_p,\Theta_p)$.
Note that the full conditional of $z_i$, given by (\ref{eq:full_cond_z}), is independent of $Z_{-i}$; hence
supremum or infimum over $Z_{-i}$ are not necessary.
By enforcing bounds on $(\Pi_p,\Theta_p)$ the infimum and the supremum in (\ref{eq:lower}) and (\ref{eq:upper})
can be made to be bounded away from 0 and 1 for points whose distribution functional values {\it a priori}
were bounded away from 0 and 1. Also, (\ref{eq:lower}) and (\ref{eq:upper}) will
take the values 0 or 1 for the points which had, respectively, distributional functional values 0 or 1 {\it a priori}.
In other words, there is a single set of points receiving positive masses
under the probability mass functions associated with both (\ref{eq:lower}) and (\ref{eq:upper}), which we subsequently prove
to be distribution functions. This set is also exactly the same set of points receiving positive masses
under the probability mass function corresponding to the distribution function {\it a priori}.

Thus, the
support of $\Theta_p$ would be compact, and that of $\Pi_p$ would be a compact subset
of its original support.
This is not an unrealistic assumption since in all practical situations, parameters
are essentially bounded away from the extreme values. In fact,
the prior on the parameters is expected to contain at least
the information regarding the range of the parameters. In almost all practical
applications, this range is finite, which, in principle, is possible to elicit. We believe that
non-compact parameter spaces are assumed
only due to the associated analytic advantages (for instance, generally integrals are easier
to evaluate analytically under the full support) and because
of the difficulty involved in elicitation of proper priors with truncated support.
However, we show in Section \ref{subsec:2comp_results},
that truncation of the support of $\Pi_p$ need not always be necessary.

In order to decide upon some adequate compact support, a pilot Gibbs sampling
run with unbounded $\Theta_p$ may be implemented first, and then the effective range
of the posterior of $\Theta_p$ can be chosen as the compact support of the prior
of $\Theta_p$. This range may be further refined by subsequently running a Gibbs sampler
with the chosen compact support, comparing the resultant density with that corresponding
to the unbounded support, and, if necessary, modifying the chosen range so that the densities agree with
each other as closely as possible.
It is demonstrated with simulated examples in Sections \ref{subsec:2comp_results},
\ref{subsec:perfect_max2comp}, and with three real applications in Sections \ref{subsec:perfect_galaxy},
\ref{subsec:perfect_acidity}, and \ref{subsec:perfect_enzyme} that often
the posterior with unbounded support is almost the same as that with
compact support, obtained from pilot Gibbs sampling.
Unless otherwise mentioned, throughout
we assume compact support of $\Theta_p$.
We remark here that the compactness assumption is not needed in the case of conjugate prior on $\Theta_p$.
In that case, $\Theta_p$ will be integrated out analytically, and hence (\ref{eq:lower}) and (\ref{eq:upper})
will not involve $\Theta_p$, thus simplifying proceedings.

Had the minimizer and the maximizer of $F_i(j\mid Y,X_{-i})$ with respect to $X_{-i}$ been constant with respect to $j$,
then, trivially, (\ref{eq:lower}) and (\ref{eq:upper}) would have been distribution functions. But this is not the case unless
$z_i$ takes on only two values with positive probability, as in the case of 2-component mixture models.
However, as shown in Section \ref{sec:distribution_function}, $F^L_i(\cdot\mid Y)$ and $F^U_i(\cdot\mid Y)$ satisfy the properties of distribution
functions for any discrete random variable. So, their inversions will sandwich all possible realizations obtained by inverting $F_i(\cdot\mid Y,X_{-i})$,
irrespective of any $X_{-i}$.

To clarify the sandwiching argument, we first define the inverse of any distribution function $F$ by
$F^{-}(x)=\inf\{y:F(y)\geq x\}$.
Further, let $R_{Z,t}=\{R_{z_i,t};i=1,\ldots,n\}$ be a common set of $iid$ random numbers
used to simulate
$Z$ at time $t$ for Markov chains starting at all possible initial values.
If we define $z_{it}={F_i}^{-}(R_{z_i,t}\mid Y,X_{-i})$, $z^L_{it}={F^U_i}^{-}(R_{z_i,t}\mid Y)$ and
$z^U_{it}={F^L_i}^{-}(R_{z_i,t}\mid Y)$, then, for all possible $X_{-i}$, it holds that
$z^L_{it}\leq z_{it}\leq z^U_{it}$ for $i=1,\ldots,n$ and $t=1,2,\ldots$.
These imply that once all $z_i$; $i=1,\ldots,n$,
drawn by inverting $F^L_i$ and $F^U_i$ coalesce, then so will every realization of $Z$ drawn from $F_i(\cdot\mid X_{-i})$, for $i=1,\ldots,n$, starting at all
possible initial values.

Analogous to $\{R_{Z,t};t=1,2,\ldots\}$, let $\{R_{\Pi_p,t}; t=1,2,\ldots\}$ and $\{R_{\Theta_p,t}; t=1,2,\ldots\}$
denote sets of $iid$ random numbers needed to generate $\Pi_p$ and $\Theta_p$, respectively,
in a hypothetical CFTP algorithm, where Markov chains from all possible starting values are simulated,
with $Z$ updated first.
Once $Z$ coalesces, so will $(\Pi_p,\Theta_p)$ since their full conditionals
(see (\ref{eq:full_cond_lambda}), (\ref{eq:full_cond_mu}) and (\ref{eq:full_cond_pi}))
show that the corresponding deterministic random mapping
function depends only upon $Z$, $\{R_{\Pi_p,t}; t=1,2,\ldots\}$, and $\{R_{\Theta_p,t}; t=1,2,\ldots\}$.
We remark here that the random numbers can always be thought of as realizations of $Uniform(0,1)$, since
the deterministic random mapping function can always be represented in terms of $Uniform(0,1)$ random numbers.

The key idea is illustrated algorithmically below.
\begin{algo}\label{algo:cftp_known}
\topline
CFTP for mixtures with known number of components
\botline
\normalfont
\ttfamily
\end{algo}
\begin{itemize}
\item[(i)]
 For $i=1,\ldots,n$, and for $\ell=1,\ldots,p$,
calculate $F^L_i(\ell\mid Y)$ and $F^U_i(\ell\mid Y)$,
given by (\ref{eq:lower}) and (\ref{eq:upper})
using some efficient optimization method. We recommend 
simulated annealing; see Section \ref{subsec:simulated_annealing}.

\item[(ii)] For $j=1\ldots$, until coalescence of $Z$, repeat steps (iii) and (iv) below.
\item[(iii)] Define $\mathcal S_j=\{-2^j+1,\ldots,-2^{j-1}\}$ for $j\geq 2$,
and let $\mathcal S_1=\{-1,0\}$. For each $m\in\mathcal S_j$,
generate random numbers $R_{Z,m}$, $R_{\Pi_p,m}$ and $R_{\Theta_p,m}$ from $Uniform(0,1)$;
once generated, treat them as fixed thereafter for all iterations.
It is important to note that at step $-2^j$ no random number generation is required since
that step can be viewed as the initializing step, where all possible chains, from all possible
initial values of $Z$, $\Pi_p$, and $\Theta_p$, are started.

\item[(iv)]
For $t=-2^j+1,\ldots,-1,0$, determine $z^L_{it}=F^{U-}_i(R_{z_{i,t}}\mid Y)$ and
$z^U_{it}=F^{L-}_i(R_{z_{i,t}}\mid Y)$ $\forall~ i=1,\ldots,n$.
This step can be thought of as initializing the perfect 
sampler with all possible values of
$Z$ and $(\Pi_p,\Theta_p)$
at step $-2^j$, and then moving on to the next forward step following
generation of $Z$ independently of the previous step, using the above random numbers
and optimized distribution functions. 
Generation of $(\Pi_p,\Theta_p)$ is not necessary because
of the sandwiching relation $z^L_{it}\leq z_{it}\leq z^U_{it}$, which holds for any $(\Pi_p,\Theta_p)$,
and because coalescence of $z^L_{it}$ and $z^U_{it}$ $\forall~ i=1,\ldots,n$
for some $t\leq 0$ guarantees coalescence of all chains corresponding to $(\Pi_p,\Theta_p)$.

\item[(v)] If $z^L_{it^*}=z^U_{it^*}$ $\forall~ i=1,\ldots,n$ and for some $t^*<0$, then
run the following Gibbs sampling steps from $t=t^*$ to $t=0$:
\begin{itemize}
\item[(a)] Let $Z^*=(z^*_1,\ldots,z^*_n)'$ denote the coalesced value of $Z$ at time $t^*$.
Given $Z^*$, draw $(\Pi^*_p,\Theta^*_p)$ from the full conditionals (\ref{eq:full_cond_pi}),
(\ref{eq:full_cond_lambda}) and (\ref{eq:full_cond_mu}) in order,
using the corresponding random numbers already generated. Thus, $(Z^*,\Pi^*_p,\Theta^*_p)$
is the coalesced value of the unknown quantities at $t=t^*$.
Importantly, it is not straightforward to sample from full conditionals of
non-conjugate distributions and/or in the case of compact parameter spaces.
In such situations we recommend rejection sampling/adaptive rejection sampling;
see Section \ref{subsec:rejection_sampling} for details.
\item[(b)] Carry forward the above Gibbs sampling chain started at $t=t^*$ till $t=0$,
simulating sequentially
from (\ref{eq:full_cond_z}), (\ref{eq:full_cond_pi}), (\ref{eq:full_cond_lambda}) and
(\ref{eq:full_cond_mu}). Then, the 
output of the Gibbs sampler obtained at $t=0$,
which we denote by $(Z_0,\Pi_{p0},\Theta_{p0})$, is a perfect sample
from the true target posterior distribution.
\end{itemize}
\end{itemize}
\rmfamily
\botline
Note that here optimization is required only once, in Step (i) of Algorithm \ref{algo:cftp_known}.
By reducing the gaps between the bounding chains, the algorithm can be made further
efficient. Such techniques are
discussed in Section \ref{subsec:efficiency_bounds} in conjunction with
Section \ref{sec:2comp_example}.
In fact, we present a variant of the above algorithm for two-component mixtures in
Algorithm \ref{algo:cftp_2comp} where optimization with respect to the mixture component
probability is not required. That algorithm exploits a monotonicity structure which
is not enjoyed by mixtures having more than two components. Indeed, even though Algorithm
\ref{algo:cftp_known} is applicable for mixtures with any known number of components,
Algorithm \ref{algo:cftp_2comp} is applicable only to two-component mixtures.

It is interesting to note that we need to run just two chains (\ref{eq:lower}) and
(\ref{eq:upper}) and check
their coalescence;
there is no need to simulate $(\Pi_p,\Theta_p)$ before coalescence occurs with respect to $Z$ in these two bounding chains, even in
non-conjugate cases.
This property of our methodology has some important advantages which are detailed in Section \ref{subsec:advantages}.

It is proved in Section \ref{sec:validity} that coalescence of $Z$ occurs almost surely in finite time. \ctn{Foss98} showed that coalescence occurs
in finite time if and only if the underlying Markov chain is uniformly ergodic. In Section \ref{sec:uniform_ergodicity} we show that our
Gibbs sampler, which first updates $Z$, is uniformly ergodic, which is expected thanks to the compact
parameter space.
The proofs in Sections \ref{sec:distribution_function}, \ref{sec:validity} and \ref{sec:uniform_ergodicity} go through with the modified bounds needed for mixtures with unknown
number of components.

In conjugate setups further simplification results since we only need to simulate perfectly
from the posterior of $Z$; once a perfect sample of $Z$ is obtained, simulations
from (\ref{eq:full_cond_pi}), (\ref{eq:full_cond_lambda}), and (\ref{eq:full_cond_mu})
ensures exact samples from the posteriors of $(\Pi_p,\Theta_p)$ as well. In order to simulate
from the posterior of $Z$, we can integrate out $(\Pi_p,\Theta_p)$ from the full conditional
of $z_i$ and construct the bounding chains with respect to the marginalized distribution function
of $z_i$, optimizing the marginalized distribution function with respect to $Z_{-i}$.

\subsection{Efficiency of the bounding chains}
\label{subsec:efficiency_bounds}

It is an important question to ask if the lower bound (\ref{eq:lower}) can be made larger or if
the upper bound (\ref{eq:upper}) can
be made smaller, to accelerate coalescence.
This can be achieved if a monotonicity structure can be identified
in $(\Pi_p,\Theta_p)$.
In Section \ref{sec:2comp_example} we illustrate this with an example.
In Section \ref{subsec:sb_bounding_chains} we propose a method
for reducing the gaps between the bounds in mixture models with unknown number
of components.
There it is also discussed that for these models, more information in the data can further reduce the gap between the bounding chains.

\subsection{Restricted parameter space and rejection sampling after coalescence}
\label{subsec:rejection_sampling}

If our algorithm coalesces at time $t<0$, then Gibbs sampling is necessary from
that point on till time $t=0$. The bounds, however, may prevent exact simulation from the
full conditionals of $\Theta_p$
using conventional methods,
such as the Box-Muller transformation (\ctn{Box58}) in the case of normal full conditionals, which
becomes truncated normal under the restrictions.
In these situations, rejection sampling may be used. Briefly, let $\{R^*_{rt};r=1,2,\ldots\}$ denote
a collection of infinite random numbers, to be used sequentially for rejection sampling of the continuous
random variables at time $t$ by the full conditionals of the continuous random variables.
Actual simulation using rejection sampling is not necessary until $Z$ coalesces.
In the case of non-conjugate priors (perhaps, in addition to restricted parameter space),
the full conditional densities are often log-concave. In such situations the same
principle can be used, but with rejection sampling replaced by adaptive rejection sampling
(\ctn{Gilks92a}, \ctn{Gilks92b}).

\subsection{Advantages of our approach}
\label{subsec:advantages}

Our bounding chain approach for only the discrete components $Z$ has several advantages over the previous approaches. Firstly,
simulation of the continuous parameters before coalescence of $Z$, is unnecessary. This advantage is important because
construction of bounds for the continuous parameters, even if possible, may not be useful since the coalescence
probability of continuous parameters corresponding to the bounding chains, is zero. Moreover, bounding the distribution functions
of continuous parameters in the mixture model context does not seem to be straightforward without discretization. Another advantage
of our perfect sampling principle is that we do not need a partial order of the multi-dimensional state space and it is unnecessary
to find minimal and maximal elements to serve as initial values of the bounding chains. Indeed, our bounding chains begin with simulations
from $F^L_1(\cdot\mid Y)$ and $F^U_1(\cdot\mid Y)$, which do not require any initial values.
Also, importantly, our approach of creating bounds for $Z$ does not depend upon the assumption of conjugate
priors. Exactly the same approach
will be used in the case of non-conjugate priors. After coalescence, regardless of
compact support or non-conjugate priors, Gibbs sampling can be carried out in a very straightforward manner
till time $t=0$.

\subsection{Obtaining infimum and supremum of $F_i(\cdot\mid Y, X_{-i})$ in practice}
\label{subsec:simulated_annealing}

For each $j=1,\ldots,p$, and for all $i=1,\ldots,n$, the bounds
$F^L_i(j\mid Y)$ and $F^U_i(j\mid Y)$
are bounded away from 0 and 1 but not always easily available
in closed forms. Numerical optimization using simulated annealing (see, for example, \ctn{Robert04} and the references therein)
with temperature $T\propto \frac{1}{\log(1+t)}$, where $t$ is the iteration number, turned out to be very effective in our case.
This is because the method, when properly tuned,
can be quite accurate, and it is entirely straightforward to handle constraints
(introduced through the restricted parameter space in our methodology) with simulated annealing
through the acceptance-rejection steps as in Metropolis-Hastings algorithm.
At each time $t$ a set of fixed random numbers will be used for implementation of simulated annealing
within our perfect sampling methodology.

Interestingly, for our perfect sampling algorithm
we do not need simulated annealing to be arbitrarily accurate; given random numbers $\{R_{Z,t};t=1,2,\ldots\}$ we only need it to be
accurate enough to generate the same realization from the approximated distribution functions as obtained had we used the exact solution.
For instance, assume that $F^L_i(j-1\mid Y)<R_{z_i,t}\leq F^L_i(j\mid Y)$, implying that $z^L_{it}=j$. Letting $\hat F^L_i$ denote the approximated distribution function,
we only need the approximation to satisfy $\hat F^L_i(j-1\mid Y)<R_{z_i,t}<\hat F^L_i(j\mid Y)$ so that $z^L_{it}=j$ even under the approximation.
This is achievable even if arbitrarily accurate approximation is not obtained.
Since in general there seems to be no way to check the error of approximation by simulated annealing,
we propose the following method. Instead of a single run of simulated annealing with a fixed run
length, one may give a few more runs, in each case increasing the length of the run
by a moderately large integer, and obtain $z^L_{it}$ and $z^U_{it}$ in each case.
Since the random numbers $R_{Z,t}$ are fixed (in fact, we recommend
fixing the random numbers used for simulated annealing as well),
the values of $z^L_{it}$ and $z^U_{it}$ will become constants as
the run length increases for simulated annealing. These constants must be the same as would have been obtained
had the optimization been exact. This strategy is not excessively burdensome computationally,
since there is no need to carry out each run afresh; the output of the last iteration of the current
run of simulated annealing will be taken as the initial value for the next run.

Our perfect sampling methodology is illustrated in a 2-component normal mixture example
in Section \ref{sec:2comp_example};
here we simply note that our
method worked excellently. A further experiment associated with
the same example, and reported in Section \ref{subsec:2comp_simulated_annealing},
showed that perfect sampling based on simulated annealing yielded results exactly
the same as those obtained by perfect sampling based on exact optimization, in 100\% of 100,000 cases.
The outcome of the latter experiment clearly encourages the use of simulated annealing for optimization
in perfect sampling.

We now extend our perfect sampling methodology to mixtures with
unknown number of components, which is a variable-dimensional problem. In this context, the non-parametric approach of \ctn{Escobar95} and the reversible jump
MCMC (RJMCMC) approach of \ctn{Richardson97} are pioneering. The former uses Dirichlet process (see, for example, \ctn{Ferguson73})
to implicitly induce variability in the number of components, while maintaining a fixed-dimensional framework, while the latter directly treats the number
of components as unknown, dealing directly, in the process, with a variable dimensional framework.
The complexities involved with the latter framework makes it difficult to extend our perfect sampling methodology
to the case of RJMCMC.  A new, flexible mixture model based on Dirichlet process has been introduced by \ctn{Bhattacharya08} (henceforth, SB), which is
shown by \ctn{Sabya10a} (see also \ctn{Sabya10}) to include \ctn{Escobar95} as a special case,
and is much more efficient and computationally cheap
compared to the latter.
Hence, we develop a perfect sampling methodology for the model of SB, which automatically applies to \ctn{Escobar95}.

\section{Perfect sampling for normal mixtures with unknown number of components}
\label{sec:sb_mixture}

As before, let $Y=(y_1,\ldots,y_n)'$ denote the available data set. SB considers the following model
\begin{equation}
[y_i\mid\Theta_M]\sim \frac{1}{M}\sum_{j=1}^M \sqrt{\frac{\lambda_j}{2\pi}} \exp\left\{-\frac{\lambda_j}{2}(y_i-\mu_j)^2\right\}
\label{eq:normixture_sb}
\end{equation}
In the above, $M$ is the maximum number of components the mixture can possibly have, and is known;
$\Theta_M$ = $\{\theta_1, \theta_2, \ldots, \theta_M\}$ with
$\theta_j$ = ($\mu_j, \lambda_j$), where $\lambda_j=\sigma^{-2}_j$.
We further assume that $\Theta_M$ are samples drawn from a Dirichlet process:
\begin{eqnarray}
\theta_j&\stackrel{iid}{\sim}& G\nonumber\\
G&\sim& DP(\alpha G_0)\label{eq:G_0}
\end{eqnarray}
Usually a $Gamma$ prior is assigned to the scale parameter $\alpha$.

Under the mean distribution $G_0$ in (\ref{eq:G_0}),
\begin{eqnarray}
\lambda_j &\stackrel{iid}{\sim}& Gamma \left(\frac{\eta}{2},\frac{\zeta}{2}\right)\label{eq:sb_prior_lambda} \\
\left[\mu_j\mid \lambda_j\right] &{\sim}& N(\mu_0, \psi\lambda^{-1}_j)\label{eq:sb_prior_mu}
\end{eqnarray}

Under the Dirichlet process assumption the parameters $\theta_j$ are coincident with positive probability; because of this (\ref{eq:normixture_sb}) reduces to the form
\begin{equation}
[y_i\mid\Theta_M]=\sum_{j=1}^p \pi_j \sqrt{\frac{\lambda^*_j}{2\pi}} \exp\left\{-\frac{\lambda^*_j}{2}(y_i-\mu^*_j)^2\right\} ,
\label{eq:normixture2_sb}
\end{equation}
where $\left\{\theta^*_1,\ldots,\theta^*_p\right\}$ are $p$ distinct components in $\Theta_M$ with $\theta^*_j$ occurring $M_j$ times, and $\pi_j=M_j/M$.

Using allocation variables $Z=(z_1,\ldots,z_n)'$, SB's model can be represented as follows:
For $i=1,\ldots,n$ and $j=1,\ldots,M$,
\begin{eqnarray}
\left[y_i\mid z_i=j,\Theta_M\right]&=&\sqrt{\frac{\lambda_j}{2\pi}} \exp\left\{-\frac{\lambda_j}{2}(y_i-\mu_j        )^2\right\}
\label{eq:y_given_z}\\
\left[z_i=j\right]&=&\frac{1}{M}
\label{eq:latent_z}
\end{eqnarray}
As is easily seen and is argued in \ctn{Sabya10}, setting $M=n$ and $z_i=i$ for $i=1,\ldots,M (=n)$,
that is, treating $Z=(1,2,\ldots,n)'$ as non-random, yields the Dirichlet process mixture model of \ctn{Escobar95}.

However, unlike the case of mixtures with fixed number of components, the full conditionals of only $Z$ and $\Theta_M$
can not be used to construct an efficient perfect sampling algorithm in the case of unknown number of components.
This is because the full conditional of $\theta_j$ given the rest depends upon
$Z$ as well as $\Theta_{-jM}$, which implies that even if $Z$ coalesces, $\theta_j$ can not coalesce
unless $\Theta_{-jM}$ also coalesces. But this has very little probability of happening in one step.
Of more concern is the fact that $Z$ may again become non-coalescent if $\Theta_M$ does not coalesce
immediately after $Z$ coalesces. Hence, although the algorithm will ultimately converge, it may take
too many iterations.
This problem can be bypassed by considering the reparameterized version of the model, based on the distinct elements of $\Theta_M$ and
the configuration indicators.

\subsection{Reparameterization using configuration indicators and associated full conditionals}
\label{subsec:reparameterization}

As before we define the set of allocation variables $Z=(z_1,\ldots,z_n)'$, where $z_i=j$ if $y_i$
is from the $j$-th component.
Letting $\Theta^*_M=\{\theta^*_1,\ldots,\theta^*_k\}$ denote the distinct components in $\Theta_M$, the
element $c_j$ of the configuration vector $C=(c_1,\ldots,c_M)'$ is defined as $c_j=\ell$ if and only if
$\theta_j=\theta^*_{\ell}$; $j=1,\ldots,M$, $\ell=1,\ldots,k$. Thus, $(Z,\Theta_M)$ is reparameterized
to $(Z,C,k,\Theta^*_M)$, $k$ denoting the number of distinct components in $\Theta_M$.

The full conditional distribution of $z_i$ is given by
\begin{equation}
[z_i=j\mid Y,C,k,\Theta^*_M]\propto
\sqrt{\frac{\lambda_j}{2\pi}}\exp\left\{-\frac{\lambda_j}{2}(y_i-\mu_j)^2\right\}
\label{eq:sb_full_cond_z}
\end{equation}
Since $\Theta_M$ can be obtained from $C$ and $\Theta^*_M$, we represented the right hand side of (\ref{eq:sb_full_cond_z})
in terms of $\Theta_M$.

To obtain the full conditional of $c_j$, first let $k_{j}$ denote the number of distinct values in $\Theta_{-jM}$, and let
$\theta^{j^*}_{\ell}$; $\ell=1,\ldots,k_{j}$ denote
the distinct values. Also suppose that $\theta^{j^*}_{\ell}$ occurs $M_{\ell j}$ times.

Then the conditional distribution of $c_j$ is given by
	  \begin{equation}
	  [c_j=\ell\mid Y,Z,C_{-j},k_j,\Theta^*_M]=\left\{\begin{array}{c}\kappa q^*_{\ell j}\hspace{2mm}\mbox{if}\hspace{2mm}\ell=1,\ldots,k_j\\ \kappa q_{0j}\hspace{2mm}\mbox{if}\hspace{2mm}\ell=k_j+1\end{array}\right.
	  \label{eq:config_fullcond_sb}
	  \end{equation}
	  where
\begin{eqnarray}
q_{0j}&=&\alpha\frac{(\frac{\zeta}{2})^{\frac{\eta}{2}}}{\Gamma(\frac{\eta}{2})}\times\left(\frac{1}{n_j\psi+1}\right)^{\frac{1}{2}}\times\left(\frac{1}{2\pi}\right)^{\frac{n_j}{2}}\nonumber\\
          &\times& \ \ \frac{2^{\frac{\eta+n_j}{2}}\Gamma(\frac{\eta+n_j}{2})}{\left\{\zeta+\frac{n_j(\bar y_j-\mu_0)^2}{n_j\psi+1}+\sum_{i:z_i=j}(y_i-\bar y_j)^2\right\}^{\frac{\eta+n_j}{2}}},\label{eq:sb_q0}\\
	  q^*_{\ell j}&=&M_{\ell j}\frac{(\lambda^{j^*}_{\ell})^{\frac{n_j}{2}}}{(2\pi)^{\frac{n_j}{2}}}\exp\left[-\frac{\lambda^{j^*}_{\ell}}{2}\left\{n_{j}(\mu^{j^*}_{\ell}-\bar y_{j})^2+\sum_{i:z_i=j}(y_i-\bar y_{j})^2\right\}\right]\nonumber\\
          \label{eq:sb_ql_star}
	  \end{eqnarray}
In (\ref{eq:config_fullcond_sb}), (\ref{eq:sb_q0}), and (\ref{eq:sb_ql_star}), $\kappa$ is the normalizing constant,
$n_j=\#\{i:z_i=j\}$ and $\bar y_j=\sum_{i:z_i=j}y_i/n_j$.
Note that $q_{0j}$ is the normalizing constant of the distribution $G_j$ defined by the following:
\begin{eqnarray}
\left[\lambda_j\right] &\sim& Gamma\left(\frac{\eta+n_j}{2},\frac{1}{2}\left\{\zeta+\frac{n_j(\bar y_j-\mu_0)^2}{n_j\psi+1}+\sum_{i:z_i=j}(y_i-\bar y_j)^2\right\}\right)\nonumber\\
\label{eq:lambda_base_measure}\\
\left[\mu_j\mid\lambda_j\right] &\sim& N\left(\frac{n_j\bar y_j\psi+\mu_0}{n_j\psi+1},\frac{\psi}{\lambda_j(n_j\psi+1)}\right)\label{eq:mu_base_measure}
\end{eqnarray}

The conditional posterior distribution of $\theta^*_{\ell}$ is given by
\begin{equation}
\left[\theta^*_{\ell}\mid Y,Z,C\right]\sim Gamma\left(\lambda^*_{\ell}:\eta^*_{\ell},\zeta^*_{\ell}\right)\times N\left(\mu^*_{\ell}:\mu^*_{0\ell},\psi^*_{\ell}{\lambda^*_{\ell}}^{-1}\right),
\label{eq:sb_theta_star}
\end{equation}
where
\begin{eqnarray}
n^*_{\ell}&=&\sum_{j:c_j=\ell}n_j,\hspace{3mm}\bar y^*_{\ell}=\sum_{j:c_j=\ell}n_j\bar y_j\Big/\sum_{j:c_j=\ell}n_j,\hspace{3mm}\eta^*_{\ell}=\frac{n^*_{\ell}+\eta}{2},\label{eq:notation1}\\
\mu^*_{0\ell}&=&\left(\psi n^*_{\ell}\bar y^*_{\ell}+\mu_0\right)/\left(\psi n^*_{\ell}+1\right),\label{eq:notation2}\\
\psi^*_{\ell}&=&\psi\big / \left(\psi n^*_{\ell}+1\right),\label{eq:notation3}
\end{eqnarray}
and
\begin{eqnarray}
\zeta^*_{\ell}&=&\frac{1}{2}\left\{\zeta+\frac{n^*_{\ell}(\mu_0-\bar y^*_{\ell})^2}
{\psi n^*_{\ell}+1} + \sum_{j:c_j=\ell}n_j(\bar y_j-\bar y^*_{\ell})^2 +\sum_{j:c_j=\ell}\sum_{i:z_i=j}(y_i-\bar y_j)^2\right\}.\label{eq:notation4}
\end{eqnarray}
It is to be noted that the $\theta^*_{\ell}$ are conditionally independent.

For Gibbs sampling, we first update $Z$, followed by updating $C$ and the number of distinct components $k$,
and finally  $\{\theta^*_{\ell};\ell=1,\ldots,k\}$.

\subsection{Non-conjugate $G_0$}
\label{subsec:sb_non_conjugate}

In the case of non-conjugate $G_0$ (which may have the same density form as a conjugate prior but with
compact support), $q_{0j}$ is not available in closed form. We then modify our
Gibbs sampling strategy by bringing in auxiliary variables in a way similar to that of Algorithm 8  in \ctn{Neal00}.
To clarify, let $\theta^a=(\mu^a,\lambda^a)$ denote an auxiliary variable (the superscript ``$a$" stands for auxiliary).
Then, before updating $c_j$ we first simulate from the full conditional distribution of $\theta^a$ given the current $c_j$ and
the rest of the variables as follows: if $c_j=c_{\ell}$ for some $\ell\neq j$, then $\theta^a\sim G_0$. If, on
the other hand, $c_j\neq c_{\ell}$ $\forall\ell\neq j$, then we set $\theta^a=\theta^*_{c_j}$. Once $\theta^a$ is obtained
we then replace the intractable $q_{0j}$ with the tractable expression
\begin{equation}
q^a_j= \alpha\frac{(\lambda^a_{j})^{\frac{n_j}{2}}}{(2\pi)^{\frac{n_j}{2}}}\exp\left[-\frac{\lambda^a_{j}}{2}\left\{n_{j}(\mu^a_{j}-\bar y_{j})^2+\sum_{i:z_i=j}(y_i-\bar y_{j})^2\right\}\right]
 \label{eq:sb_nonconjugate_form}
 \end{equation}
 Once $c_j$ is simulated, if it is observed that $\theta_j\neq\theta^a$ $\forall j$, then $\theta^a$ is discarded.

 \subsection{Relabeling $C$}
 \label{subsec:relabeling}
 Simulation of $C$ by successively simulating from the full conditional distributions (\ref{eq:config_fullcond_sb}) incurs
 a labeling problem. For instance, it is possible that all $c_j$ are equal even though each of them corresponds to
 a distinct $\theta_j$. For an example, suppose that $\Theta^*_M$ consists of $M$ distinct elements, and $c_j=M$ $\forall j$.
 Then although there are actually $M$ distinct components, one ends up obtaining just one distinct component.
 For perfect sampling we create a
 labeling method which relabels $C$ such that the relabeled version, which we denote by $S=(s_1,\ldots,s_M)'$, coalesces if $C$ coalesces.
 To  construct $S$ we first simulate $c_j$ from (\ref{eq:config_fullcond_sb}); if $c_j\in\{1,\ldots,k_j\}$, then we set $\theta_j=\theta^*_{c_j}$
 and if $c_j=k_j+1$, we draw $\theta_j=\theta^*_{c_j}\sim G_j$. The elements of $S$ are obtained from
 the following definition of $s_j$:
 $s_j=\ell$ if and only if $\theta_j=\theta^*_{\ell}$. Note that $s_1=1$ and $1\leq s_j\leq s_{j-1}+1$.
 In Section \ref{sec:s_coalescence} it is proved that
 coalescence of $C$ implies the coalescence of $S$, irrespective of the value of $\Theta^*_M$ associated with $C$.

 \subsection{Full conditionals using $S$}
 \label{subsec:fullcond_modified}
With the introduction of $S$ it is now required to modify some of the full conditionals of the unknown random variables, in addition
to introduction of the full conditional distribution of $S$.
The form of the full conditional $[z_i\mid Y,S,k,\Theta^*_M]$ remains the same as (\ref{eq:sb_full_cond_z}), but $\Theta_M$ involved in the right hand side
of (\ref{eq:sb_full_cond_z}) is now obtained from $S$ and $\Theta^*_M$. The modified  full conditional of $c_j$, which we denote
by $[c_j\mid Y,Z,S_{-j},k_j,\Theta^*_M]$, now depends upon $S_{-j}$, rather than $C_{-j}$, the notation being clear from the context.
The form of this full conditional remains the same as (\ref{eq:config_fullcond_sb}) but now the distinct components
$\theta^{j^*}_{\ell}$; $\ell=1,\ldots,k_{j}$ are associated with the corresponding components of $S$ rather than $C$. The form of the modified full
conditional distribution of $\theta^*_{\ell}$, which we now denote by $\left[\theta^*_{\ell}\mid Y,Z,S,k\right]$, remains the same as (\ref{eq:sb_theta_star}),
but in equations (\ref{eq:notation1}) to (\ref{eq:notation4}), $C$ must be replaced by $S$. In the above full conditionals, $k$ and $k_j$ are now assumed
to be associated with $S$.

The conditional posterior $[S\mid Y,C,\Theta_M]$ gives point mass to $S^*$, where $S^*=(s^*_1,\ldots,s^*_M)'$ is the relabeling obtained from $C$ and $\Theta_M$
following the method described in Section \ref{subsec:relabeling}.
For the construction of bounds, the individual full conditionals $[s_j\mid Y, S_{-j},C,\Theta_M]$, giving full mass to $s^*_j$,
will be considered due to the
convenience of dealing with distribution functions of one variable.
It follows that once $Z$ and $C$ coalesces, $S$ and $\Theta^*_M$ must also coalesce.
In the next section we describe how to construct efficient bounding chains for $Z$, $C$ and $S$.
Bounding chains for $S$ are not strictly necessary as it is possible to optimize the bounds for $Z$ and $C$ with respect to $S$, but the efficiency of the other bounding chains
is improved, leading to an improved perfect sampling algorithm,
if we also construct bounding chains for $S$.

\subsection{Bounding chains}
\label{subsec:sb_bounding_chains}

As in the case of mixtures with known number of components, here also
the idea of constructing bounding chains is associated with distribution functions of the discrete random variates,
but here the bounding chains can be made efficient by fixing the already coalesced individual discrete variates
while taking the supremum and the infimum of the distribution functions.
Moreover, for informative data, the full conditional distributions of $c_j$ (hence, of $s_j$) will be similar given any values of the conditioned
variables; thus the difference between the supremum and the infimum of their distribution functions are expected to be small.
This particular heuristic is reflected in the results of the application of our methodology to three real data sets in Section \ref{sec:real_application}.
Also, as noted in Section \ref{subsec:bound_known_k}, even in the case of unknown number of components, $\Theta^*_M$ can
be analytically marginalized out in conjugate cases, simplifying optimization procedures.
The full conditional distributions associated with our model, marginalized over $\Theta^*_M$ in a conjugate case
are provided in \ctn{Sabya10a}.

\subsubsection{Bounds for $Z$}
\label{subsubsec:z_bound}

Let $F_{z_i}(\cdot\mid Y,S,k,\Theta^*_M)$ denote
the distribution function of the full conditional of $z_i$, and let $F_{c_j}(\cdot\mid Y, S_{-j},k_j,\Theta^*_M)$,
$F_{s_j}(\cdot\mid Y, S_{-j},C,\Theta_M)$ stand for those of $c_j$ and $s_j$, respectively. Also assume that
$-\infty<M_1\leq\mu_j\leq M_2<\infty$ and $0\leq M_3\leq\lambda_j\leq M_4<\infty$, for all $j$.

Let $\bar S$ denote the set consisting of only those $s_j$ that have coalesced, and let $S^-=S\backslash\bar S$ consist of the
remaining $s_j$. Then
\begin{eqnarray}
F^L_{z_i}\left(\cdot\mid Y,\bar S\right)&=&\inf_{S^-,k,\Theta^*_M}F_{z_i}(\cdot\mid Y,\bar S,S^-,k,\Theta^*_M)\label{eq:inf_z}\\
F^U_{z_i}\left(\cdot\mid Y,\bar S\right)&=&\sup_{S^-,k,\Theta^*_M}F_{z_i}(\cdot\mid Y,\bar S,S^-,k,\Theta^*_M)\label{eq:sup_z}
\end{eqnarray}
Clearly, fixing $\bar S$ helps reduce the gap between (\ref{eq:inf_z}) and (\ref{eq:sup_z}).
The infimum and the supremum above can be calculated by simulated annealing. For the proposal mechanism needed
for simulated annealing with $\bar S$ held fixed, we selected
$s_j\in S^-$ uniformly from $\{1,\ldots,s_{j-1}+1\}$, where $s_{j-1}$ either belongs to $\bar S$ or has been selected uniformly
from $\{1,\ldots,s_{j-2}+1\}$. Once $S$ is proposed in this way, this determines $k$ automatically. We then propose $\theta^*_1,\ldots,\theta^*_k$
using normal random walk proposals with approximately optimized variance.

\subsubsection{Bounds for $C$}
\label{subsubsec:c_bound}
Let $\bar Z$ denote the set of coalesced $z_i$, and let $Z^-=Z\backslash\bar Z$ consist of those $z_j$ that did not yet coalesce.
Then
\begin{eqnarray}
F^L_{c_j}\left(\cdot\mid Y,\bar S, \bar Z\right)&=&\inf_{S^-,k_j,Z^-,\Theta^*_M}F_{c_j}(\cdot\mid Y,\bar S,S^-,k_j,\bar Z,Z^-,\Theta^*_M)\label{eq:inf_c}\\
F^U_{c_j}\left(\cdot\mid Y,\bar S, \bar Z\right)&=&\sup_{S^-,k_j,Z^-,\Theta^*_M}F_{c_j}(\cdot\mid Y,\bar S,S^-,k_j,\bar Z,Z^-,\Theta^*_M)\label{eq:sup_c}
\end{eqnarray}
Note that for $\bar S=\emptyset$, the supremum corresponds to $k_j=1$ and the infimum corresponds to
$k_j=M-1$. If $\bar S\neq\emptyset$, the supremum is associated with $k_j=\#\bar S\backslash\{s_j\}$, the number of distinct components of $\bar S\backslash\{s_j\}$,
and the infimum corresponds to the case where $k_j=\# \left(\bar S\cup S^-\right)\backslash\{s_j\}$, when all elements of $S^-$ are distinct.
Thus, proposal mechanism of $S^-$ for simulated annealing is not necessary; manually setting all elements of $S^-$ to be equal for obtaining the supremum
and manually setting all elements of $S^-$ to be distinct for obtaining the infimum are sufficient, since the actual values of the elements of $S$ are unimportant.
This strategy dictates the number of distinct elements of $\Theta_M$, and their positions.
The proposal mechanism
for $\Theta^*_M$ may be chosen to be the same as that used for obtaining the bounds for $z_i$, while the elements of $Z^-$ may be proposed
by drawing uniformly from $\{1,\ldots,M\}$.

\subsubsection{Bounds for $S$}
\label{subsubsec:s_bound}

Letting $\bar C$ and $C^-=C\backslash\bar C$ denote the sets of coalesced and the non-coalesced $c_j$, the lower
and the upper bounds for the distribution function of $s_j$ are
\begin{eqnarray}
F^L_{s_j}\left(\cdot\mid Y,\bar C\right)&=&\inf_{C^-,\Theta^*_M}F_{s_j}(\cdot\mid Y,\bar C,C^-,\Theta^*_M)\label{eq:inf_s}\\
F^U_{s_j}\left(\cdot\mid Y,\bar C \right)&=&\sup_{C^-,\Theta^*_M}F_{s_j}(\cdot\mid Y,\bar C,C^-,\Theta^*_M)\label{eq:sup_s}
\end{eqnarray}
For simplicity let us denote $F_{s_j}(\cdot\mid Y,\bar C,C^-,\Theta^*_M)$
by $F_{s_j}(\cdot)$ suppressing the conditioned variables.
Since, given $C$ and $\Theta^*_M$, $S$ is uniquely determined, $F_{s_j}(k)=0$ or 1, for $k = 1 ,\ldots ,M$.
Thus, optimization of $F_{s_j}(k)$ needs to be carried out extremely carefully because either the correct optimum or the
incorrect optimum will be obtained, leaving no scope for approximation.
However, simulated annealing is unlikely to perform adequately in this situation. For instance, while maximizing, a long
sequence of iterations yielding
$F_{s_j}(k)=0$ does not imply that 1 is not the maximum. Similarly, a long sequence of 1's while minimizing may mislead
one to believe that 1 is the minimum.
In other words, the algorithm does not exhibit gradual move towards the optimum, making
convergence assessment very difficult.
So, we propose to construct functions $h_j(\cdot)$ of
$F_{s_j}(\cdot)$'s and appropriate
auxiliary variables such that the optimization of $F_{s_j}(\cdot)$ is embedded in the optimization of $h_j(\cdot)$, while
avoiding the aforementioned problems by allowing gradual move towards the optimum.
Details are provided below.

\subsubsection*{A more convenient optimizing function}
We construct $h_j(\cdot)$ as follows:
\begin{equation}
h_j(W,F) = \sum_{i=1}^{M} w_{i}\left\{\frac{F_{s_j}(i)+w_i}{1+w_{i}}\right\}^{\frac{1}{2}}
\label{eq:h}
\end{equation}
where $W$ = $(w_1,\ldots,w_M)$ denotes the vector of weights, $F$ = $(F_{s_j}(1),\ldots,F_{s_j}(M))$ and
$\sum_{j=1}^{M} w_j = 1$ with $w_j > 0, \forall j$.
Clearly, $0<h_j(\cdot)<1$.
We represent
$w_j$ as $w_j = \frac{n_j}{\sum_{i=1}^{M}n_i}$, where $n_i>0$.
We use simulated annealing to optimize (\ref{eq:h}) with respect to $(W,C^-,\Theta^*_M)$ but
let $n_k \rightarrow \infty$ with the iteration number while simulating other $n_i;i\neq k$ randomly from some bounded interval.
This leads to optimization of $F_{s_j}(k)$, while avoiding the problems of naive simulated annealing.
In our examples we took $n_k\propto\log(1+t)$, where $t$ is the iteration number.

\subsubsection*{Optimizing strategy}

Since $S$ is just a relabeled version of $C$, the distribution functions of the full conditionals of $c_j$ and $s_j$ are
optimized by the same $\Theta_M$,
provided that none of $s_j$ coalesced during optimization in the case of $C$.
All that the proposal mechanism
requires then is to simulate $c_j\in C^-$ uniformly from $\{1,\ldots,M\}$. If $C$ ($=\bar C\cup C^-$) and $\Theta_M$ do not lead to
a valid $S$, then the proposal is to be rejected, remaining at the current $C^-$, else the acceptance-rejection step of simulated annealing
is to be implemented. If, on the other hand, some $s_j$ had coalesced during optimization in $c_j$, the optimizer in the case of $s_j$ is expected
to be a slight modification of that in the case of $c_j$.
We construct the modification as follows.
If $C$, simulated from the bounding chains (\ref{eq:inf_c}) and (\ref{eq:sup_c}) in the previous step,
is not compatible with $\Theta_M$,
then we augment $\Theta^*_M$ with new components drawn uniformly: $\mu \sim U(M_1,M_2)$ and
$\lambda \sim U(M_3,M_4)$, in such a manner that compatibility is ensured.
We then use the adjusted set of $\Theta_M$ for rest of the annealing steps. This scheme worked adequately in all our
experiments.
Note that if entire $C$ coalesces, then for all $j$ and for any $\Theta_M$ associated with $C$,
$F^L_{s_j}\left(\cdot\mid Y,\bar C \right)=F^U_{s_j}\left(\cdot\mid Y,\bar C \right)=F_{s_j}(\cdot\mid Y,C,\Theta_M)$, which implies
coalescence of $S$ (recall the discussion in Section \ref{subsec:fullcond_modified}).

The proof presented in Section \ref{sec:distribution_function} goes through to show that the bounds of the distribution functions of $(Z,C,S)$, which are obtained by
optimizing the original functions
treating the coalesced random variates as fixed, are also distribution functions.
The proof remains valid even if the original distribution functions of the discrete variates are optimized with respect to the scale parameter $\alpha$
and other hyper-parameters. Optimization with respect to the latter is necessary if $\alpha$ and the hyper-parameters are treated as unknowns and must be simulated
perfectly, likewise as $\Theta_M$.
Assuming that the original Gibbs sampling algorithm is updated by first updating $Z$,
then $C$, followed by $S$, and finally $\Theta^*_M$, the proof of coalescence of the random variables in finite time is exactly as that provided in Section \ref{sec:validity}.
The proof of uniform ergodicity presented in Section \ref{sec:uniform_ergodicity} applies with minor modifications in the current
mixture problem with unknown number
of components.

Below we provide an algorithmic representation of perfect sampling in
mixtures with unknown number of components.
\begin{algo}\label{algo:cftp_unknown}\topline
CFTP for mixtures with unknown number of components \botline \normalfont \ttfamily
\end{algo}
\begin{itemize}
\item[(1)] For $j=1\ldots$, until coalescence of $(Z,C)$, repeat steps (2) and (3) below.
\item[(2)] Define $\mathcal S_j=\{-2^j+1,\ldots,-2^{j-1}\}$ for $j\geq 2$,
and let $\mathcal S_1=\{-1,0\}$. 
For each $m\in\mathcal S_j$,
generate random numbers $R_{Z,m}$, $R_{C,m}$, $R_{S,m}$ and $R_{\Theta_M,m}$ (again, 
we shall
let these random numbers stand for realizations from the uniform distribution on $(0,1)$), meant
for simulating $Z$, $C$, $S$, and $\Theta_M$ respectively. Although random numbers
are not necessary for simulating $S$ from its optimized full conditionals because of degeneracy,
$R_{S,m}$ will still be used for optimizing its distribution function using simulated annealing.
The random numbers $R_{\Theta_M,m}$ will correspond to $M$ distinct components, so that
the same set will suffice for smaller numbers of distinct components in the set $\Theta_M$
where all components need not be distinct. The distinct components of $\Theta_M$ are meant to be simulated
(but recall that actual simulation is not necessary until the coalescence of $(Z,C,S)$)
using those random numbers in the set $R_{\Theta_M,m}$, which correspond to their positions in $\Theta_M$.

Once generated, treat the random numbers as fixed thereafter for all iterations.
As in Algorithm \ref{algo:cftp_known}, at step $-2^j$ no random number generation is required.
\item[(3)] For $t=-2^j+1,\ldots,-1,0$, implement steps (3) (i), (3) (ii) and (3) (iii):
\begin{itemize}
\item[(i)] For $i=1,\ldots,n$,
\begin{itemize}
\item[(a)] For $\ell=1,\ldots,M$, calculate $F^L_{z_i}(\ell\mid Y,\bar S)$ and
$F^U_{z_i}(\ell\mid Y,\bar S)$, using the simulated annealing
method described in Section \ref{subsubsec:z_bound}.
\item[(b)] Determine $z^L_{it}=F^{U-}_{z_i}(R_{z{i,t}}\mid Y,\bar S)$ and
$z^U_{it}=F^{L-}_{z_i}(R_{z{i,t}}\mid Y,\bar S)$.
As in Algorithm \ref{algo:cftp_known}, this step can be thought of as initializing
the perfect sampler with all possible values of $(Z,C,S,k,\Theta^*_M)$
at step $-2^j$; at step $-2^j+1$, $\bar S=\emptyset$ (omitting the always coalescent $s_1=1$),
signifying that simulations at this forward step is independent of the previous step $-2^j$.
From step $-2^j+2$ onwards, there is positive probability that $\bar S\neq\emptyset$.
Thus, with positive probability, the bounding chains for $z_i$ will be more efficient
from this point on.
\end{itemize}
\item[(ii)] For $i=1,\ldots,M$,
\begin{itemize}
\item[(a)] For $\ell=1,\ldots,k_i+1$, calculate $F^L_{c_i}(\ell\mid Y,\bar S,\bar Z)$ and
$F^U_{c_i}(\ell\mid Y,\bar S,\bar Z)$, using the simulated annealing technique
described in Section \ref{subsubsec:c_bound}. Recall that the supremum corresponds to $k_i=\#\bar S\backslash\{s_i\}$, when $S^-$ contains
a single distinct element,
and the infimum corresponds to the case where $k_i=\# \left(\bar S\cup S^-\right)\backslash\{s_i\}$, when all elements of $S^-$ are distinct, and so
the set $S^-$ will be set manually to have a single distinct element or all distinct elements.

\item[(b)] Set $c^L_{it}=F^{U-}_{c_i}(R_{c_{i,t}}\mid Y,\bar S,\bar Z)$ and
$c^L_{it}=F^{U-}_{c_i}(R_{c_{i,t}}\mid Y,\bar S,\bar Z)$.
\end{itemize}

\item[(iii)] For $i=1,\ldots,M$,
\begin{itemize}
\item[(a)] For $\ell=1,\ldots,M$, calculate $F^L_{s_i}(\ell\mid Y,\bar C)$ and
$F^U_{s_i}(\ell\mid Y,\bar C)$, using the methods described in Section \ref{subsubsec:s_bound}.

\item[(b)]
Since, for some $\ell^*\in\{1,\ldots,M\}$,
$F^L_{s_i}(\ell\mid Y,\bar C)=0$ for $\ell<\ell^*$ and 1 for $\ell\geq \ell^*$,
it follows that $s^L_{it}=\ell^*$. Similarly, $s^U_{it}$ can be determined.
\end{itemize}

\end{itemize}

\item[(4)] If, for some $t^*<0$, $z^L_{it^*}=z^U_{it^*}$ $\forall i=1,\ldots,n$, and $c^L_{it^*}=c^U_{it^*}$
$\forall i=1,\ldots,M$, then
run the following Gibbs sampling steps from $t=t^*$ to $t=0$:
\begin{itemize}
\item[(a)] Let $Z^*=(z^*_1,\ldots,z^*_n)'$ and $C^*=(c^*_1,\ldots,c^*_M)'$ denote the coalesced values of $Z$
and $C$ respectively, at time $t^*$.
Given $(Z^*,C^*)$, arbitrarily choose any value of $\Theta_M$ which is compatible with $C^*$
(one way to ensure compatibility is to choose any $\Theta_M$ having $M$ distinct elements); then
obtain $S^*$ from $[S\mid\bY,C,\Theta_M]$ using the algorithm given in Section \ref{subsec:relabeling}.
Finally, obtain $\Theta^*_M$ from its full conditional distribution, using the random numbers
already generated.
As in Algorithm \ref{algo:cftp_known}, here also rejection sampling/adaptive rejection sampling
may be necessary for obtaining $\Theta^*_M$.
This yields the coalesced value $(Z^*,C^*,S^*,\Theta^*_M)$ at time $t=t^*$.

\item[(b)] Using the random numbers already generated, carry forward the above Gibbs sampling
chain started at $t=t^*$ till $t=0$,
simulating, in order, from the full conditionals of $(Z,C,S,\Theta^*_M)$, provided in
Sections \ref{subsec:reparameterization}, \ref{subsec:sb_non_conjugate},
\ref{subsec:relabeling}, and \ref{subsec:fullcond_modified}.
Then, the output of the Gibbs sampler obtained at $t=0$,
which we denote by $(Z_0,C_0,S_0,\Theta^*_{M0})$, is a perfect sample
from the true target
 posterior distribution.

\end{itemize}
\end{itemize}
\rmfamily
\botline

\subsection{Illustration of perfect simulation in a mixture with maximum two components}
\label{subsec:perfect_max2comp}

We illustrate our new methodologies in the framework of the mixture model of SB assuming $M=2$.
In other words, we consider the model
\begin{equation}
[y_i\mid \Theta_2] \sim  \frac{1}{2}\sum_{j=1}^{2}N(y_i;\mu_j, \lambda^{-1}_j)
\label{eq:sb_2comp}
\end{equation}
We further assume that $\lambda_1=\lambda_2=\lambda$, where $\lambda$ is assumed to be known. Hence,
$\Theta_2=(\theta_1, \theta_2)$, where $\theta_j=\mu_j$, $j=1,2$.
As in the case of the two-component mixture example detailed in Section \ref{sec:2comp_example},
here also we consider a simplified model for convenience of illustration and to validate the reliability of
simulated annealing as the optimizing method in our case.

We specify the prior of $\mu_j$ as follows:
\begin{eqnarray}
\mu_j&\stackrel{iid}{\sim}& G,\hspace{2mm}j=1, 2\nonumber\\
 G &\sim& \mathcal{D}(\alpha G_0),\nonumber\\
\end{eqnarray}
and
$\mu_j\stackrel{iid}{\sim} N(\mu_0, \psi\lambda^{-1})$ under $G_0$.

We draw 3 observations $y_1,y_2,y_3$, from (\ref{eq:sb_2comp}) after fixing $\mu_1=2.19$, $\mu_2=2.73$ and $\lambda=20$.
We chose $\alpha=1$, $\mu_0=1.98$, and $\psi=2.33$ (the latter two are drawn from normal
and inverse gamma distributions). Using a pilot Gibbs sampling run we set $0.45=M_1\leq \mu_1,\mu_2\leq M_2=3.7$.

\subsubsection{Optimizer for bounding the distribution function of $z_i$}
\label{subsubsec:z_opt}

The exact minimizer and the maximizer of the distribution function of $z_i$ with respect to $\Theta_2$ or the reparameterized variables $(S,\Theta^*_2)$ are of the form $(a,b)$ where each of $a$ and $b$ can take the
values $y_i$, $M_1$ or $M_2$. Evaluation of the distribution function at these points yields the desired minimum and the maximum at different
time points $t$.

\subsubsection{Optimizer for bounding the distribution function of $c_j$}
\label{subsubsec:c_opt}
For $c_j$, the optimizer with respect to $\Theta_2$ is given by $(a,b)$ where $a$ and $b$ can take the values $\bar y_j$, $M_1$ and $M_2$.
Of course, this is the same as what would be obtained by optimizing with respect to the reparameterized version $(S,\Theta^*_2)$.
As before,
evaluation of the distribution function at these points is necessary for obtaining the desired optimizer. In this case, the optimizer with respect to $Z$ is obtained
by considering all possible values of $Z=(z_1,z_2,z_3)'$.

\subsubsection{Optimizer for bounding the distribution function of $s_j$}
\label{subsubsec:s_opt}

No explicit optimization is necessary to obtain the bounds for $s_j$, as $S=(s_1,s_2)$ is completely determined by $C$ obtained from its corresponding bounding chains.
Note that for the four possible values of $C=(c_1,c_2)$: $(1,1)$, $(1,2)$, $(2,1)$, $(2,2)$, the corresponding values of $S=(s_1,s_2)$ are $(1,1)$, $(1,2)$, $(1,1)$ and $(1,2)$, respectively.

\subsubsection{Results of perfect sampling}
\label{subsubsec:results_perfect_unknown}

\begin{figure}
\begin{center}
\includegraphics[width=11cm,height=9cm]{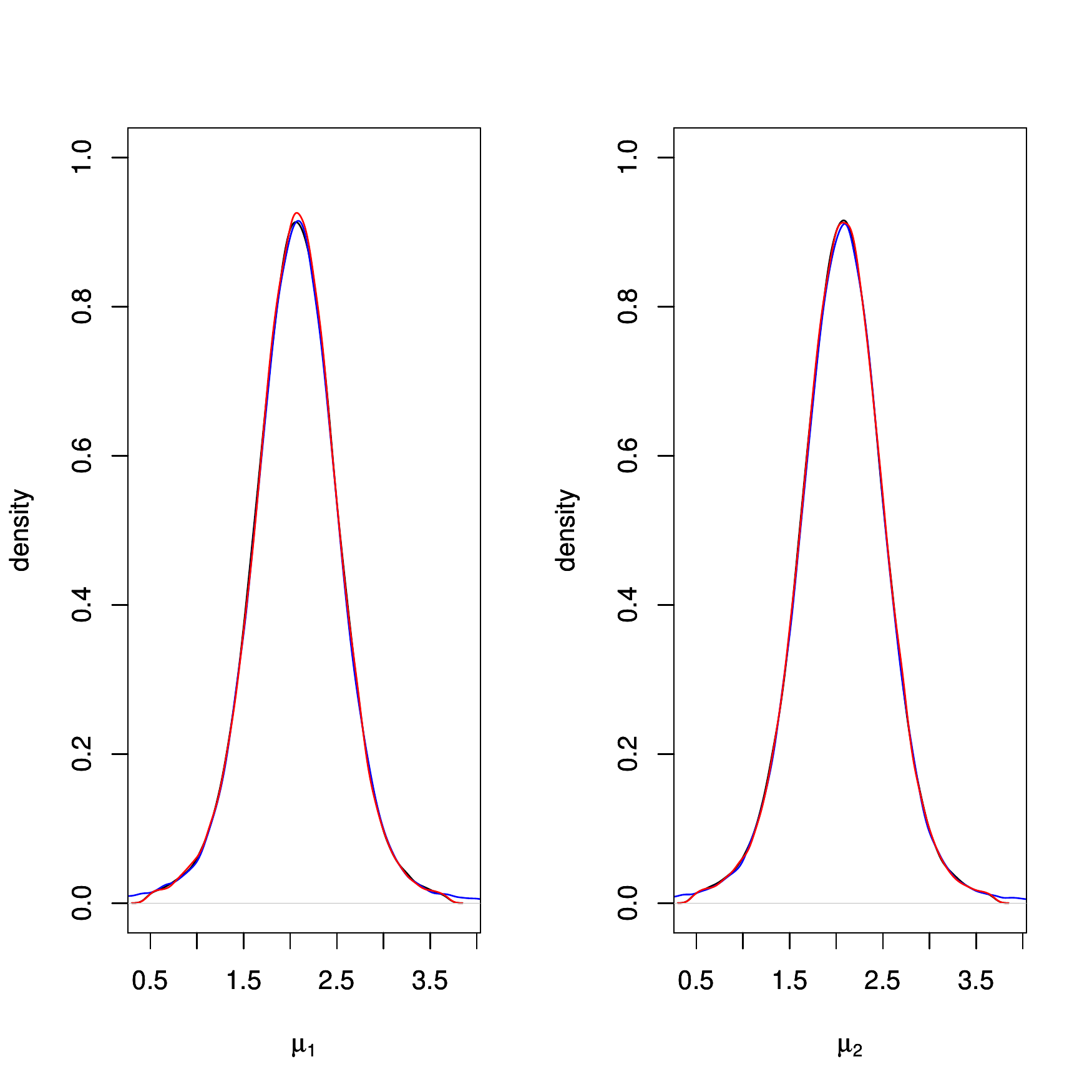}
\caption{Posterior densities of $\mu_1$ and $\mu_2$ using samples obtained from perfect simulation
(red curve) and independent runs of Gibbs sampling (black curve). The blue curve stands for the posteriors
corresponding to the unbounded support.}
\label{fig:compare_dpp}
\end{center}
\end{figure}

Results of $100,000$ $iid$ perfect samples are displayed in Figure \ref{fig:compare_dpp}; the results are compared with $100,000$ independent
Gibbs sampling runs, each time discarding the samples obtained in the first $10,000$ Gibbs sampling iterations and retaining only the sample in the
$10,001$-th iteration. The figure shows that the posterior distributions corresponding to
perfect sampling (red curve), Gibbs sampling (black curve) in the case of compact supports,
as well as the posterior corresponding to unbounded support (again, based on 100,000 Gibbs samples,
each with a burn-in of length 10,000), displayed by the blue curve, agree with each other very closely.
This is very encouraging, and validates our perfect sampling methodology.
\begin{figure}
\begin{center}
\includegraphics[width=11cm,height=9cm]{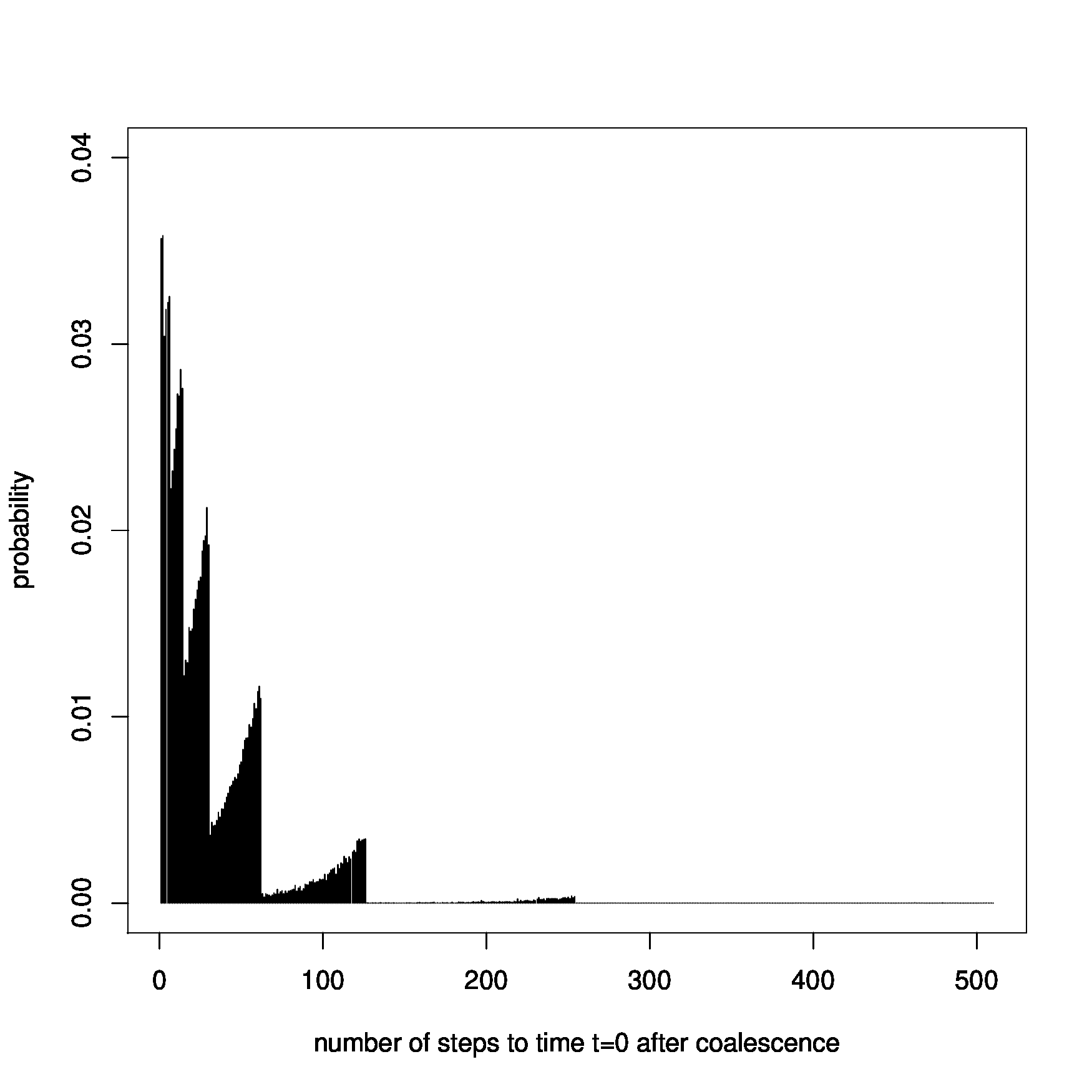}
\caption{Probability distribution of the number of steps taken from the time of coalescence till the time $t=0$
corresponding to Figure \ref{fig:compare_dpp}, associated with 100,000 $iid$ perfect samples.}
\label{fig:steps}
\end{center}
\end{figure}

It is important to remark in this context of validation that a reviewer had expressed concern
that the period between the time of coalescence till the time $t=0$ in our perfect
sampling algorithm may be long enough to be regarded as a burn-in period, thus rendering our $iid$ perfect
samples anyway similar to the $iid$ Gibbs samples, thus decreasing the strength of the validation exercise.
We assure the reader, however, that this is not the case.
Indeed, the effective maximum time period between the coalsescence time and the time $t=0$
in the probability distribution of the period between the coalescence time
and the time $t=0$, displayed in Figure \ref{fig:steps}, in all 100,000 cases
is 254, which also has very small probability of occurrence (only 35 out of 100,000).
The number of occurrences of the values between 255 and 506 (the latter being the actual maximum
time period between the coalsescence time and the time $t=0$) is either 0 or 1 (mostly zero), in all 100,000 cases.
In fact, the modal time period between the coalescence time and the time $t=0$ is just
2, occurring 3582 times. In other words, in most cases it took just two steps to reach time
$t=0$ after coalescence. As a result, the period between the coalescence time and the time $t=0$ is just
too small to be any reasonable burn-in period,
and is not comparable to the burn-in period of length 10,000 of our Gibbs sampler.

The above arguments show that the agreement between perfect sampling and Gibbs sampling in
Figure \ref{fig:compare_dpp} is due
to the fact that both the algorithms are correct, the former being exact, and the latter being approximate, but
very accurate thanks to the long burn-in of length 10,000.

\subsubsection{Validation of simulated annealing in this example}
\label{subsubsec:validation_simulated_annealing}

As in the example with known number of components here also we validate simulated annealing by separately obtaining $100,000$ $iid$ samples
using our perfect sampling algorithm but using simulated annealing (with 7,000 iterations) to optimize the bounds for the distribution
functions of $(Z,C,S)$.
We have used the same
random numbers as used in the perfect sampling experiment for obtaining $100,000$ $iid$ samples using the exact bounds.
All the corresponding samples
at time $t=0$ turned out to be the same, just as in the example of the mixture with exactly two components. This
obviously encourages the use of simulated annealing
in perfect sampling from mixtures with unknown number of components.

\section{Application of perfect simulation to real data}
\label{sec:real_application}

We now consider application of our perfect sampling methodology to three real data sets---Galaxy, Acidity,
and Enzyme data. Both RG and SB used all the three data sets to illustrate their methodologies.
The Galaxy data set consists of 82 univariate observations on velocities of galaxies, diverging from our own galaxy.
The second data concerns an acidity index measured in a sample of
155 lakes in north-central Wisconsin.
The third data set concerns the distribution of enzymic activity in the blood,
for an enzyme involved in the metabolism of carcinogenic substances, among
a group of 245 unrelated individuals.

\subsection{Perfect sampling for Galaxy data}
\label{subsec:perfect_galaxy}

\subsubsection{Determination of appropriate ranges of the parameters}

We implemented a Gibbs sampler with
$M = 10$, $\eta = 4$; $\zeta = 1$; $\mu_0 = 20$; $a_{\alpha} = 10$; $b_{\alpha} = 0.5$; $\psi = 33.3$;
and obtained results quite similar to that reported in SB, who used $M=30$.
Using the results obtained in our experiments, we set the following bounds on the parameters: for $j=1,\ldots,M (=10)$,
$9.5\leq \mu_j\leq 34.5$, $0.01\leq\lambda_j\leq 5$ and $0.08\leq\alpha\leq 35.5$.
The fit to the data obtained with this set up turned out to be similar to that obtained by SB.

\subsubsection{Computational issues}

We implemented our perfect sampling algorithm with the above-mentioned hyperparameter values and parameter ranges. Our experiments
suggested that 500 simulated annealing iterations for each optimization step are adequate, since further increasing the number
of iterations did not significantly improve the optima.
The terminal chains coalesced after 32,768 steps.
The reason for the coalescence of the bounding
chains after a relatively large number of iterations may perhaps be attributed to the inadequate amount of information contained
in the relatively sparse 82-point data set required to reduce the gap between the bounding chains (recall the discussion in
Section \ref{subsec:sb_bounding_chains}). In fact, as it will be seen, perfect sampling with the other two data sets containing
much more data points and showing comparatively much clear evidence of bimodality (particularly the Acidity data set)
coalesced in much less number of steps.
However, compared to the number of steps needed to achieve coalescence,
the computation time needed to implement the steps turned out to be more serious.
In this Galaxy data, with $M=10$, the computation time taken by a workstation to implement 32,768 backward iterations turned
out to be about 11 days! We discuss in Section \ref{sec:discussion} that parallel computing is an effective way
to drastically reduce computation time.
However, we consider another experiment with $M=5$ that took just 13 hours
for implementation, yielding results very similar to those with $M=10$.

\subsubsection{Results of implementation}
\begin{figure}
\begin{center}
\includegraphics[width=11cm,height=9cm]{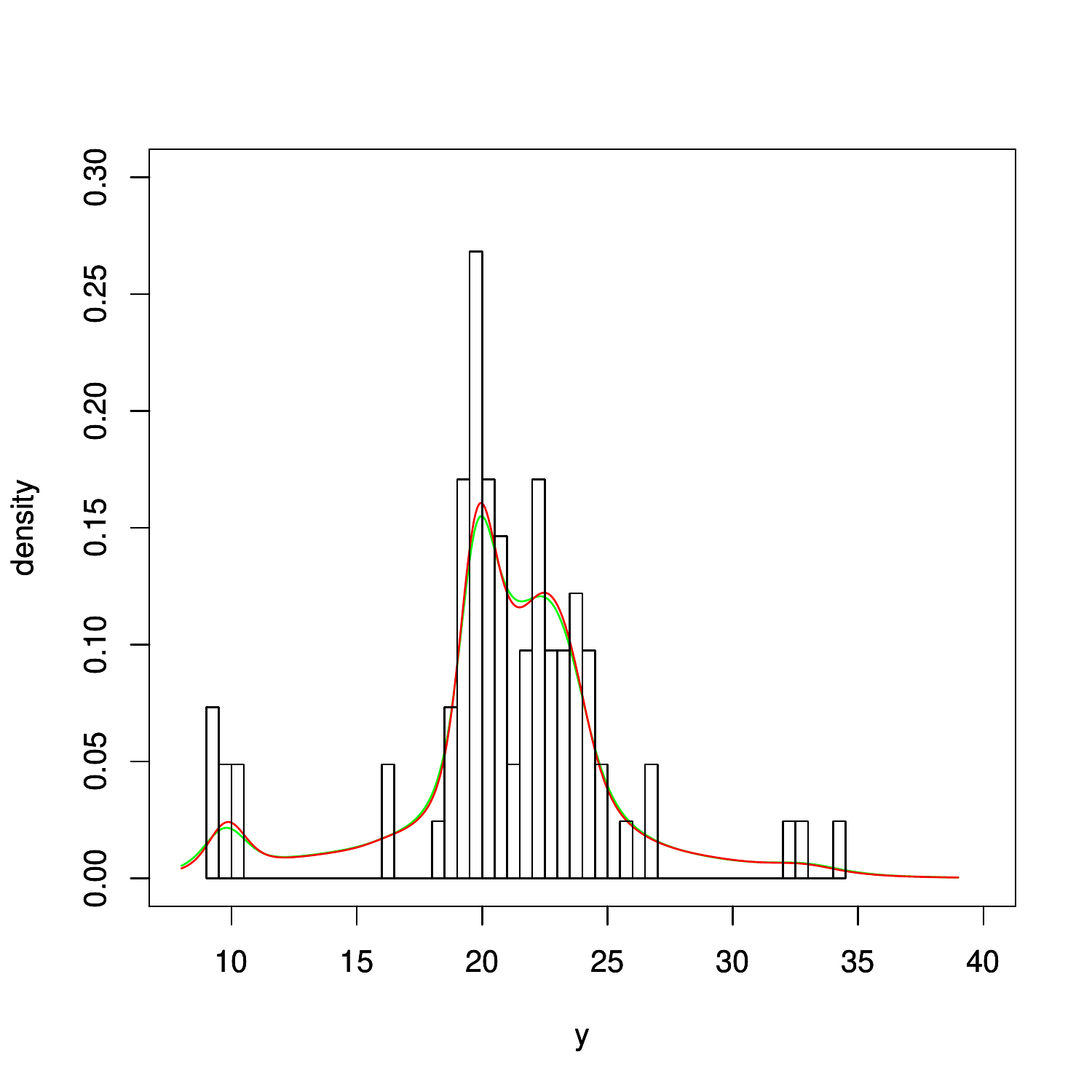}
\caption{Histogram of the Galaxy data and the posterior predictive density corresponding to
perfect simulation with $M = 10$ (red curve).
The green curve stands for the Gibbs sampling based
posterior predictive density assuming unbounded parameter space.}
\label{fig:compare_post10}
\end{center}
\begin{center}
\includegraphics[width=11cm,height=9cm]{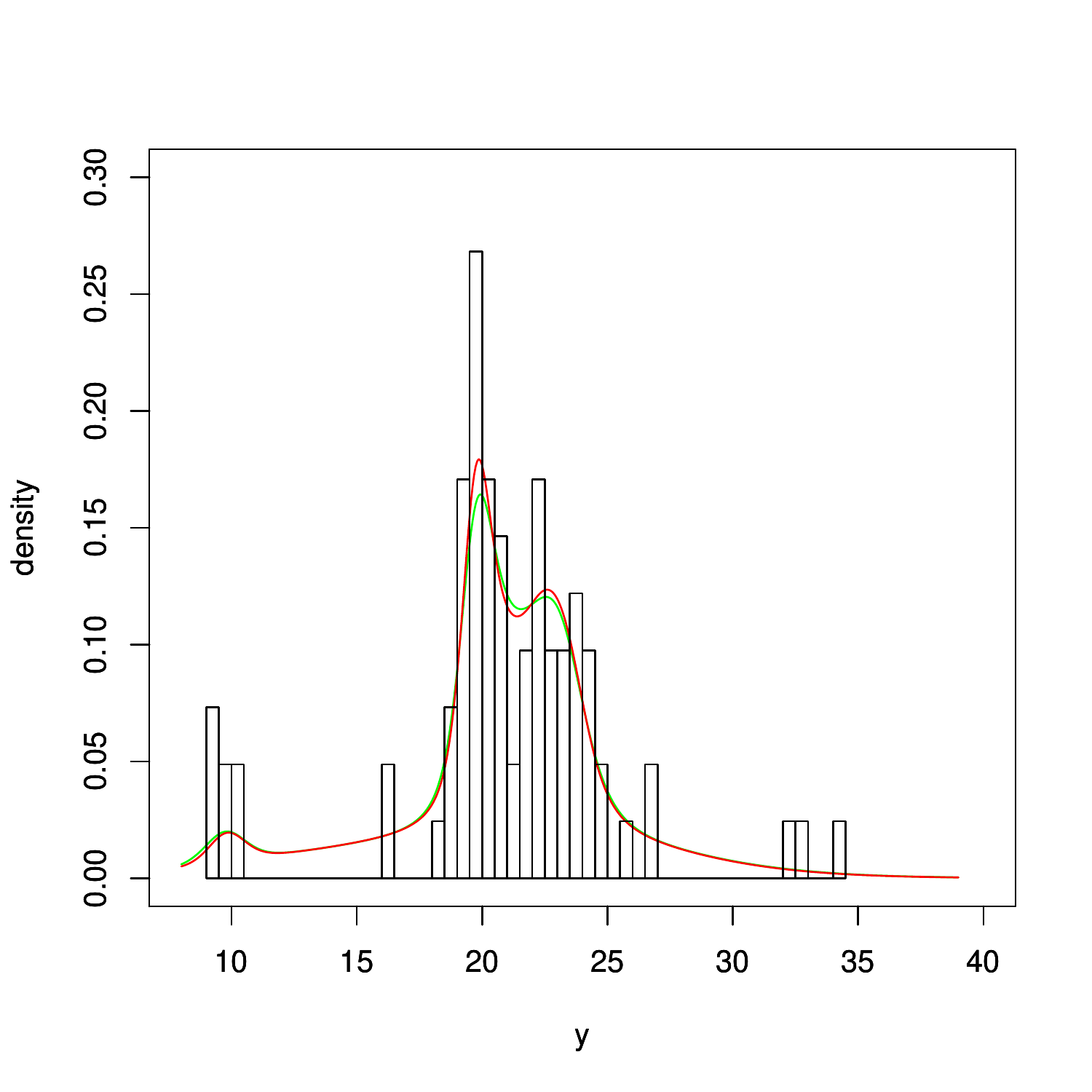}
\caption{Histogram of the Galaxy data and the posterior predictive density corresponding to perfect simulation
with $M = 5$ (red curve). The green curve stands for the Gibbs sampling based
posterior predictive density assuming unbounded parameter space.}
\label{fig:compare_post}
\end{center}
\end{figure}

After coalescence,
we ran the chain forward to time $t=0$, thus obtaining a perfect sample. We then further
generated 15,000 samples using the forward Gibbs sampler.
The red curve 
in Figure \ref{fig:compare_post10} stands for the posterior predictive density,
and the overlapped green curve is the
the Gibbs sampling based
posterior predictive density
corresponding to the unbounded parameter space. The figure shows that the difference
between the posterior predictive distributions
with respect to bounded and unbounded parameter spaces are negligible, and can perhaps be attributed to
Monte Carlo error only.
The posterior probabilities of the number of distinct components being $\{1,\ldots,10\}$ turned out to be
$\{$0, 0, 0.000067, 0.0014, 0.0098, 0.044133, 0.1358, 0.265133, 0.3436, 0.200067$\}$, respectively.

\subsubsection{Experiment to reduce computation time by setting $M=5$}
\label{subsubsec:smaller_M}

As a possible alternative to reduce computation time, 
we decided to further reduce the value of $M$ to 5.
The ranges of the parameters when $M=5$ turned out to be somewhat larger compared to the case of $M = 10$:
for $j=1,\ldots,5$, $9.5\leq \mu_j\leq 34.5$, $0.01\leq\lambda_j\leq 20$ and $0.08\leq\alpha\leq 100$.
Now the two terminal chains coalesced in 2048 steps taking about
13 hours. As before, once the terminal chains coalesced,
we ran the chain forward to time $t=0$, and then further
generated 15,000 samples using the forward Gibbs sampler.
The posterior predictive density is shown in
Figure \ref{fig:compare_post}.
As before, the figure shows that the differences between the
posterior predictive densities
with respect to bounded and unbounded parameter spaces are negligible enough to be attributed
to Monte Carlo error.
Moreover, when compared to Figure \ref{fig:compare_post10}, Figure \ref{fig:compare_post}
indicates that the fitted DP-based mixture model with $M=5$
is not much worse than that with $M=10$.
Here the posterior probabilities of the number of distinct components being $\{1,2,3,4,5\}$, respectively,
are $\{$0.000067, 0.001467, 0.026667,0.229733, 0.742067$\}$.

\subsection{Perfect sampling for Acidity data}
\label{subsec:perfect_acidity}

Following the procedure detailed in Section \ref{subsec:perfect_galaxy}
we set the following bounds: for $j=1,\ldots,M (=10)$, $4\leq\mu_j\leq 6.9$,
$0.08\leq \lambda_j\leq 25$, and $0.08\leq\alpha\leq 50$.
We implemented our perfect sampler with these ranges, and with hyperparameters $\eta=4$, $\zeta=0.7$, $\mu_0=5.02$,
$a_{\alpha}=15$, $b_{\alpha}=0.5$, and $\psi=33.3$.
As in the Galaxy data, here also 500 iterations of simulated annealing for each optimization
step turned out to be sufficient.
The terminal chains took about 4 hours to coalesce in 128 steps.

\begin{figure}
\begin{center}
\includegraphics[width=11cm,height=9cm]{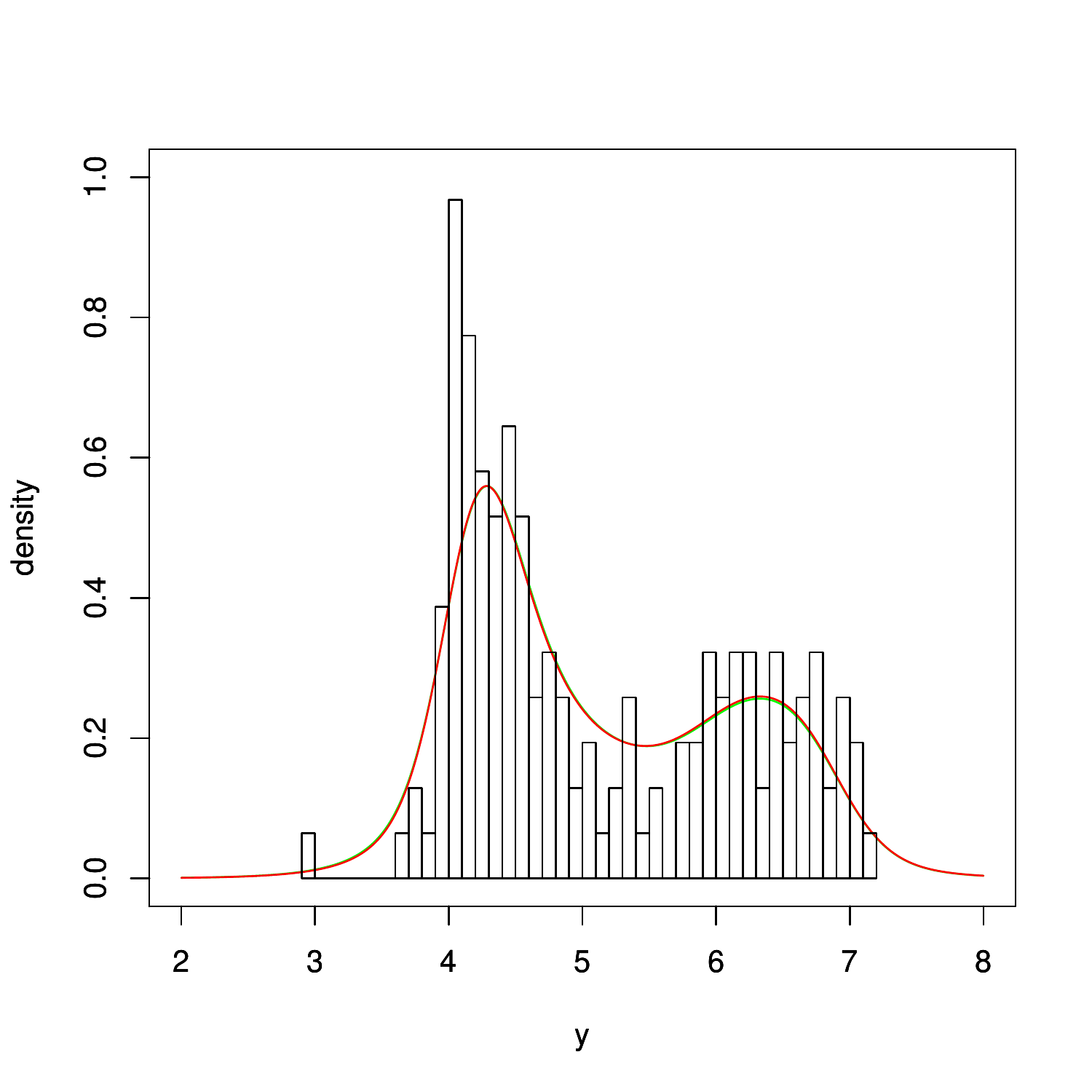}
\caption{Histogram of the Acidity data and the posterior predictive density corresponding to
perfect simulation with $M = 10$ (red curve).
The green curve stands for the Gibbs sampling based
posterior predictive density assuming unbounded parameter space.}
\label{fig:acid_post}
\end{center}
\end{figure}

The posterior predictive distribution
is shown in Figure \ref{fig:acid_post}.
Again, as before, the figure demonstrates that the posterior predictive
density remains virtually unchanged whether or not the parameter space is truncated.
Figure \ref{fig:acid_post} also
indicates that the posterior predictive distribution matches closely with that of the histogram of the data.
The posterior probabilities of the number of distinct components being $\{1,\ldots,10\}$ are
$\{$0, 0, 0.000067, 0.0024, 0.012, 0.0556, 0.159867, 0.303133, 0.323067, 0.143867$\}$, respectively.

\subsection{Perfect sampling for Enzyme data}
\label{subsec:perfect_enzyme}

Following the procedures detailed in Sections \ref{subsec:perfect_galaxy} and \ref{subsec:perfect_acidity}
we fix $M= 10$; the bounds on the parameters are: for $j=1,\ldots,M (=10)$, $0.15\leq\mu_j\leq 3$, $0.08\leq\lambda_j\leq 150.5$
and $0.08\leq\alpha\leq 50$. The hyperparameters in this example are given by
$\eta = 4$; $\zeta = 0.33$; $\mu_0 = 1.45$;
$a_{\alpha} = 20$; $b_{\alpha} = 0.5$ and $\psi = 33.3$.

\begin{figure}
\begin{center}
\includegraphics[width=11cm,height=9cm]{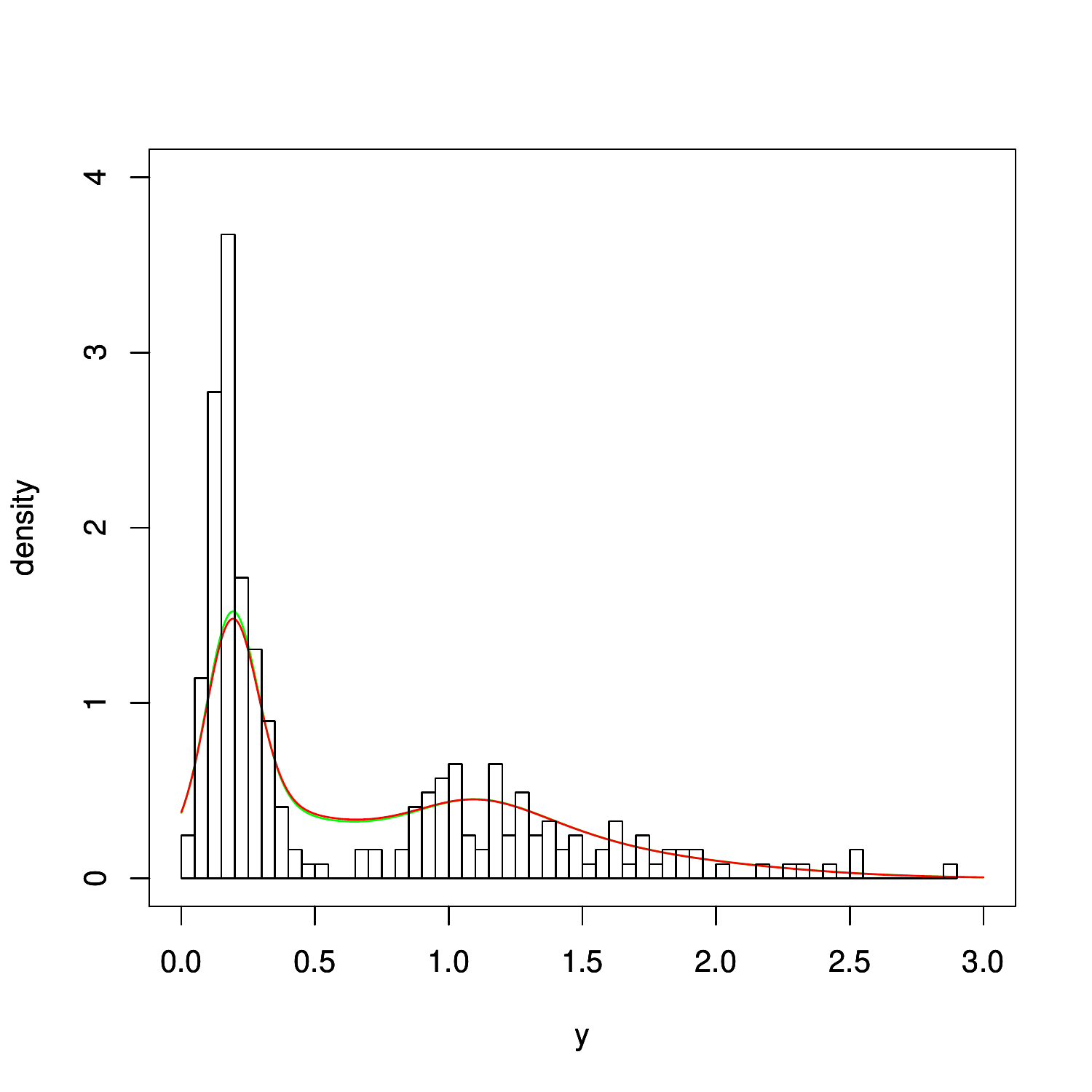}
\caption{Histogram of the Enzyme data and the posterior predictive density corresponding to
perfect simulation with $M = 10$ (red curve).
The green curve stands for the Gibbs sampling based
posterior predictive density assuming unbounded parameter space.}
\label{fig:enzyme_post}
\end{center}
\end{figure}

We implemented our perfect sampler with these specifications, along with 500 iterations of simulated
annealing for each optimization step.
The terminal chains coalesced in 2048 steps taking about 4 days.
As to be expected from the previous applications,
here also, as shown in Figure \ref{fig:enzyme_post}, truncation of the parameter space virtually makes
no difference to the resulting posterior predictive density
associated with unbounded parameter space. Good fit of the model to the data is also indicated.
The posterior probabilities of the number of distinct components being $\{1,\ldots,10\}$, respectively, are
$\{$0, 0.000933, 0.012067, 0.0634, 0.179, 0.2782, 0.219867, 0.1454,
0.075333, 0.0258$\}$.

\section{Summary, discussion and future work}
\label{sec:discussion}

We have proposed a novel perfect sampling methodology that works for mixtures where the number of components
are either known or unknown, and the set-up is either conjugate or non-conjugate. We have first developed
the method for mixtures with known number
of components, then extended it to the more important case of mixtures with unknown number of components.
Our methodology hinges upon exploiting the full conditional distributions of the discrete random variables of the problem,
optimizing the corresponding distribution functions with respect to the conditioned random variables, obtaining
upper and lower bounds of the corresponding Gibbs samplers. One particularly intriguing aspect of this
strategy is perhaps the fact that even though perfect samples of continuous random variables will also be generated,
simulation of the latter is not at all required
before coalescence of the discrete bounding chains.
We have shown that the gaps between the upper
and the lower bounds of the Gibbs sampler can be narrowed, making way for fast coalescence. Further advantages
over the existing perfect sampling procedures are also discussed in detail.
It is also easy to see that our current methodology need not be confined to univariate data, and
the same methodology goes through for handling multivariate instances.

With simulation studies we have validated our methodology for mixtures with known, as well as with unknown,
number of components.
However, application to real data sets revealed substantial computational burden, and obtaining a single perfect sample
took several hours with our limited computational resources. Thus, even though the convergence (burn-in)
issue is completely eliminated, obtaining
$iid$ realizations from the posteriors turned out to be infeasible.
As discussed in Section \ref{subsec:perfect_galaxy}, the difficulties
are likely to persist in problems where large values of the maximum number of components are plausible,
and in sparse data sets. Computational challenges are also likely to
appear in massive data sets, since then the number of allocation variables for perfect sampling will increase
manyfold. In multivariate data sets too, the computation can be excessively burdensome---here
the number of discrete simulations necessary remains the same as in the corresponding univariate problem,
but optimization with respect to the continuous variables may be computationally expensive because
of increased dimensionality.
In such situations, parallel computing can be of great help. Indeed, in a parallel computing environment
the upper and lower bounding chains can be simulated in different parallel processors, which would greatly reduce
the computation time.
Moreover, quite importantly, $iid$ simulations from the posteriors can also be carried out easily by simulating perfect samples
independently in separate parallel processors. This can be done most efficiently by utilizing two processors for each perfect realization,
so that, say, with 16 parallel processors 8 perfect $iid$ realizations can be obtained in about half the time a single perfect realization is generated in
a stand-alone machine. The parallel computing procedure can be repeated to obtain as many $iid$ realizations as desired within a reasonable
time. Increasing the number of parallel processors can obviously speed up this procedure many times, which would make implementation
of our algorithm routine.
Although we, the authors, have the expertise in parallel computing, we are yet to have access
to parallel computing facilities, which is the reason why we could not obtain perfect $iid$ realizations in our real
data experiments and could not experiment with large $M$ or massive data.
In the near future, however, such access is expected, and then it will be easier for us to elaborate on these
computational issues.

\section*{Acknowledgements}
Our sincere gratitude goes to the Editor-in-Chief, an anonymous Editor,
an anonymous Associate Editor, and two anonymous
referees, whose comments have led to an improved version of our manuscript.

\renewcommand\thefigure{S-\arabic{figure}}
\renewcommand\thetable{S-\arabic{table}}
\renewcommand\thesection{S-\arabic{section}}

\begin{center}
{\bf\LARGE Supplementary Material}
\end{center}

Throughout, we refer to our main manuscript as MB.

\section{Proof that ${F^L}_i$ and ${F^U}_i$ are distribution functions}
\label{sec:distribution_function}

Letting $X_{-i}$ denote all unknown variables other than $z_i$
we need to show that for almost all $X_{-i}$ the following holds:
\begin{itemize}
\item[(i)] $\lim_{h\rightarrow -\infty}F^L_i(h)=\lim_{h\rightarrow -\infty}F^U_i(h)=0$.
\item[(ii)] $\lim_{h\rightarrow \infty}F^L_i(h)=\lim_{h\rightarrow \infty}F^U_i(h)=1$.
\item[(iii)] For any $x_1\geq x_2$, $F^L_i(x_1)\geq F^L_i(x_2)$ and $F^U_i(x_1)\geq F^U_i(x_2)$.
\item[(iv)] $\lim_{h\rightarrow x+}F^L_i(h)=F^L_i(x)$ and $\lim_{h\rightarrow x+}F^U_i(h)=F^U_i(x)$.
\end{itemize}

{\bf Proof:} 
Let $X_{-i}$ denote all unknown variables other than $z_i$. To prove (i), note that for all $h<1$, 
$F_i(h\mid X_{-i})=0$ for almost all $X_{-i}$. Hence, by (\ref{eq:full_cond_z}) of MB
and by definition,
both $F^L_i(h)$ and $F^U_i(h)$ are 0 with probability 1. Hence, $\lim_{h\rightarrow -\infty}F^L_i(h)=\lim_{h\rightarrow -\infty}F^U_i(h)=0$
almost surely.

To prove (ii) note that for all $h>p$, $F_i(h\mid X_{-i})=1$ for almost all $X_{-i}$. Hence, for $h>p$, $F^L_i(h)=F^U_i(h)=1$,
that is, $\lim_{h\rightarrow \infty}F^L_i(h)=\lim_{h\rightarrow \infty}F^U_i(h)=1$ for almost all $X_{-i}$.

To show (iii), let $h_1>h_2$. Then, since $F_i(\cdot\mid X_{-i})$ is a distribution function satisfying monotonicity, it holds that 
$F^L_i(h_2)=\inf_{X_{-i}}F_i(h_2\mid X_{-i})\leq F_i(h_2\mid X_{-i})\leq F_i(h_1\mid X_{-i})$ for almost all $X_{-i}$.
Hence, $F^L_i(h_2)\leq \inf_{X_{-i}}F_i(h_1\mid X_{-i})=F^L_i(h_1)$. 
Similarly, $F^U_i(h_1)=\sup_{X_{-i}}F_i(h_1\mid X_{-i})\geq F_i(h_1\mid X_{-i})\geq F_i(h_2\mid X_{-i})$ for almost all $X_{-i}$. Hence, 
$F^U_i(h_1)\geq\sup_{X_{-i}}F_i(h_2\mid X_{-i})=F^U_i(h_2)$.

To prove (iv), first observe that due to the monotonicity property (iii), the following hold for any $x$:
\begin{eqnarray}
\lim_{h\rightarrow x+}F^L_i(h)&\geq& F^L_i(x)\label{eq:mono1}\\
\lim_{h\rightarrow x+}F^U_i(h)&\geq& F^U_i(x)\label{eq:mono2}
\end{eqnarray}
Then observe that, due to discreteness, $F_i(\cdot\mid X_{-i})$ is constant in the interval $[x,x+\delta)$ for some $\delta>0$.
Since the supports of $F^L_i$, $F^U_i$ and $F_i(\cdot\mid X_{-i})$ for almost all $X_{-i}$ are same, $F^L_i$ and $F^U_i$
must also be constants in $[x,x+\delta)$. This implies that equality holds in (\ref{eq:mono1}) and (\ref{eq:mono2}).

Hence, both $F^L_i$ and $F^U_i$ satisfy all the properties of distribution functions.
\\[2mm]
{\bf Remark:} The right continuity property formalized by (iv) not be true for continuous variables.
Suppose $X\sim U(0, \theta)$, $\theta>0$. Here the distribution function is
$F(x\mid \theta) = \frac{x}{\theta}$, $0<x<\theta<\infty$. But
\[
\lim_{x\rightarrow 0+} \sup_{\theta}\frac{x}{\theta}\nonumber\\
= \lim_{x\rightarrow 0+} 1\nonumber\\
= 1\nonumber\\
\]
and,
\[
\sup_{\theta} \lim_{x\rightarrow 0+} \frac{x}{\theta}\nonumber\\
= \sup_{\theta} 0\nonumber\\
= 0\nonumber\\
\]
As a consequence of the above problem, attempts to construct suitable stochastic bounds for the 
continuous parameters $(\Pi_p,\Theta_p)$ may not be fruitful. In our case such problem does not arise
since we only need to construct bounds for the discrete random variables to achieve our goal.

\section{Proof of validity of our CFTP algorithm}
\label{sec:validity}


{\bf Theorem:} The terminal chains coalesce almost surely in finite time and the value obtained at 
time $t=0$ is a a realization from the target distribution. 
\\[2mm]
{\bf Proof:}






Let $z^L_{it}$ denote the realization obtained at time $t$ by inverting 
$F^U_i$, that is, $z^L_{it}={F^U_i}^{-}(R_{z_i,t})$, where 
$\{R_{z_i,t}; i=1,\ldots,n;t=1,2,\ldots\}$ is a common set of $Uniform(0,1)$ random numbers
which are $iid$ with respect to both $i$ and $t$,
used to simulate
$Z=(z_1,\ldots,z_n)'$ at time $t$ for Markov chains starting at all possible initial values.
Similarly, let $z^U_{it}={F^L_i}^{-}(R_{z_i,t})$. 
Clearly, for any $z_{it}=F^{-}_i(R_{z_i,t}\mid X_{-i})$ started with any initial value and for any 
$X_{-i}$, $z^L_{it}\leq z_{it}\leq z^U_{it}$ for all $i$ and $t$.

For $i=1,\ldots,n$ and for $j=1,2,\ldots$, we denote by $E^j_i$ the event
\[ z^L_{i,-2^j}(-2^{j-1})=z^U_{i,-2^j}(-2^{j-1}), \] which signifies that
the terminal chains and hence the individual chains started at
$t=-2^j$ will coalesce at $t=-2^{j-1}$. 
It is important to note that both $F^L_i$ and $F^U_i$ are irreducible
which has the consequence that the probability of $E^j_i$, $P(E^j_i)>\epsilon_i>0$,
for some positive $\epsilon_i$. Since, for fixed $i$, $\{E^j_i;j=1,2,\ldots\}$ depends only upon the random numbers
$\{R_{z_i,t};t= -2^j,\ldots,-2^{j-1}\}$, $\{E^j_i;j=1,2,\ldots\}$ are independent with respect to $j$. Moreover, for fixed $j$, $E^j_i$  
depends only upon the $iid$ random numbers $\{R_{z_i,-2^j};i=1,\ldots,n\}$. Hence, $\{E^j_i;i=1,\ldots,n;j=1,2,\ldots\}$ are 
independent with respect to both $i$ and $j$.

Let $\epsilon=\min\{\epsilon_1,\ldots,\epsilon_n\}$. Then 
due to independence of $\{E^j_i;i=1,\ldots,n\}$, it follows that for $j=1,2,\ldots$,
$\bar E^j=\cap_{i=1}^nE^j_i$ are independent, and
\begin{equation}
P\left(\bar E^j\right)\geq \epsilon^n
\label{eq:indep1}
\end{equation}
The rest of the proof resembles the proof of Theorem 2 of \ctn{Casella01}. In other words,
\begin{eqnarray}
P(\mbox{No coalescence after T iterations} )&\leq & \prod_{j=1}^T\left\{1-P(\bar E^j)\right\}\label{eq:indep2}\\
&=& \left\{(1-\epsilon^n)\right\}^T\rightarrow 0\ \ \mbox{as} \ \ T\rightarrow\infty\label{eq:indep3}.
\end{eqnarray}
Thus, the probability of coalescence is 1. That the time to coalesce is almost surely finite follows
from the Borel-Cantelli lemma, exactly as in \ctn{Casella01}. 

The realization obtained at time $t=0$ after occurrence of the coalescence 
event $\bar E_j$ for some $j$ yields $Z=Z_0$ exactly from its marginal posterior
distribution. Given this $Z_0$, drawing $\Pi_{p0}$ from the full conditional distribution
(\ref{eq:full_cond_pi}) of MB 
and then drawing $\Theta_{p0}$ sequentially from (\ref{eq:full_cond_lambda}) 
and (\ref{eq:full_cond_mu}) of MB 
given $Z_0$ and $\Pi_{p0}$, yields a realization $(Z_0,\Pi_{p0},\Theta_{p0})$ exactly from
the target posterior. The proof of this exactness follows readily from the general proof (see, for example,
\ctn{Prop96}, \ctn{Casella01})
that if convergent Markov chains coalesce in a CFTP algorithm during time $t\leq 0$, then the realization obtained 
at time $t=0$ is exactly from the stationary distribution.






\section{Uniform ergodicity}
\label{sec:uniform_ergodicity}

Let $P(\cdot,\cdot)$ denote a Markov transition kernel where 
$P(x,A)$ denotes transition from the state $x$ to the set $A\in\mathcal B$, $\mathcal B$
being the associated Borel $\sigma$-algebra.
If we can show that for all $x$ in the state space the following minorization holds:
\[ P(x, A)\geq\epsilon Q(A),\hspace{2mm}A\in {\mathcal{B}}, \]
for some $0<\epsilon\leq 1$ and for some probability measure $Q(\cdot)$,
then $P(\cdot,\cdot)$ is uniformly ergodic. 

In our mixture model situation the Gibbs sampling transition kernel is
\begin{eqnarray}
&&\left[Z^{(t)},\Pi^{(t)}_p,\Theta^{(t)}_p\mid Z^{(t-1)},\Pi^{(t-1)}_p,\Theta^{(t-1)}_p\right]\nonumber\\
&& \ \ = \left[Z^{(t)}\mid\Pi^{(t-1)}_p,\Theta^{(t-1)}_p,Y\right]\left[\Pi^{(t)}_p\mid Z^{(t)},Y\right]\left[\Theta^{(t)}_p\mid Z^{(t)},\Pi^{(t)}_p,Y\right]\nonumber\\
&& \ \ \geq \left\{\inf_{\Pi^{(t-1)}_p,\Theta^{(t-1)}_p}\left[Z^{(t)}\mid\Pi^{(t-1)}_p,\Theta^{(t-1)}_p,Y\right]\right\}
\left[\Pi^{(t)}_p\mid Z^{(t)},Y\right]\left[\Theta^{(t)}_p\mid Z^{(t)},\Pi^{(t)}_p,Y\right]
\label{eq:minorization}
\end{eqnarray}
The infimum in inequality (\ref{eq:minorization}) is finite since both $\Pi^{(t-1)}_p$ and $\Theta^{(t-1)}_p$ are bounded.

Denoting the right hand side of inequality (\ref{eq:minorization}) by
$g(Z^{(t)},\Pi^{(t)}_p,\Theta^{(t)}_p)$, we put 
\begin{equation}
\epsilon = \sum_{Z} \int_{\Pi_p}\int_{\Theta_p} g(Z,\Pi_p,\Theta_p)d\Pi_pd\Theta_p>0. 
\label{eq:epsilon}
\end{equation}
Since $g(\cdot)$ is bounded above by the Gibbs transition kernel which integrates to 1, it follows
from (\ref{eq:epsilon}) that $0<\epsilon\leq 1$. 
Hence, identifying the density of the $Q$-measure as $g(\cdot)/ \epsilon$, the minorization
condition required for establishment of uniform ergodicity of our Gibbs sampling chain is seen to hold.

\section{Proof that coalescence of $C$ implies the coalescence of $S$}
\label{sec:s_coalescence}

Let $C=(c_1,\ldots,c_M)'$ be coalescent.
For convenience of illustration assume that after simulating each $c_j$, followed by drawing $\theta_j$ depending
upon the simulated value of $c_j$, the entire set $S$ is obtained from the updated set of parameters $\Theta_M$.
Note that in practice, only $s_j$ will be obtained immediately after updating $c_j$ and $\theta_j$. Let $S_{-j}=\{s_1,\ldots,s_{j-1},s_{j+1},\ldots,s_M\}$.
Then $c_{j+1}=\ell$ denotes the $\ell$-th distinct element of $S_{-j}$. If $\{1,\ldots,d_j\}$ are the distinct components in $S_{-j}$,
$d_j$ being the number of distinct components, and $\ell\leq s_j$, then $s_{j+1}=\ell$. On the other hand, if $\ell<c_{j+1}\leq d_j+1$, then $s_{j+1}=s_j+1$.

Now note that $s_1=1$, which is always coalescent. If $c_2>1$, then $s_2=2$, else $s_2=1$, for all Markov chains. Hence, $s_2$
is coalescent. If $c_3>s_2$, then $s_3=s_2+1$, else $s_3=c_3$. Since $s_2$ is coalescent,
then so is $s_3$. In general, if $c_{j+1}>s_j$, then $s_{j+1}=s_j+1$, else $s_{j+1}=c_{j+1}$. Since $s_1,\ldots,s_j$ are coalescent, hence
so is $s_{j+1}$, for $j=1,\ldots,M-1$. In other words, $S$ must coalesce if $C$ coalesces.

\section{Illustration of perfect simulation with a two-component normal mixture example}
\label{sec:2comp_example}

For $i=1,\ldots,n$, data point $y_i$ has the following distribution:
\begin{equation}
[y_i\mid \pi,\Theta_2] \sim \pi N(y_i;\mu_1, \lambda^{-1}_1)+ (1-\pi) N(y_i;\mu_2,\lambda^{-1}_2),
\end{equation}
where, for the sake of simplicity in illustration, $\lambda_1$ and $\lambda_2$ are assumed known.
The reason for considering this simplified model is two-fold. Firstly, it is easy to explain complicated methodological issues
with a simple example. Secondly, the bounds of $Z$ are available
exactly in this two-component example; the results can then be compared in the same example
with approximate bounds obtained by simulated annealing. This will validate the use of simulated annealing in our methodology.

The prior of $\mu_j$; $j=1,2$, is assumed to be of the form (\ref{eq:prior_mu}) of MB.
Fixing the true values at $\pi=0.8$, $\mu_1=2.19$ and $\mu_2=2.73$,
we draw a sample of size $n=3$ from a normal mixture where
$\sigma^2_1=\lambda^{-1}_1=0.9$, $\sigma^2_2=\lambda^{-1}_2=0.5$
are considered known. The hyperparameters are set to the following values:
$\tau_1=0.9$, $\tau_2=0.8$,
$\xi_1=2.5$ and  $\xi_2=3.5$.
We illustrate our methodology in drawing samples exactly from the posterior
$[\pi,\mu_1,\mu_2\mid y_1,y_2,y_3]$.

\subsection{Construction of bounding chains}
\label{subsec:bounding_chains}

To obtain $F^L_i$ and $F^U_i$; $i=1,2,3$, note that
here we only need to minimize and maximize
\begin{equation}
F_i(1\mid X_{-i})=\frac{\pi\sqrt{\lambda_1}\exp\left\{-\frac{\lambda_1}{2}(y_i-\mu_1)^2\right\}}{\pi\sqrt{\lambda_1}\exp\left\{-\frac{\lambda_1}{2}(y_i-\mu_1)^2\right\}
+ (1-\pi)\lambda_2\exp\left\{-\frac{\lambda_2}{2}(y_i-\mu_2)^2\right\}}
\label{eq:2comp_cdf}
\end{equation}
with respect to $\mu_1$, $\mu_2$ and $\pi$.
Based on a pilot Gibbs sampling run we obtain the following bounds for $\mu_1$ and $\mu_2$:
$M_1=0.2\leq\mu_1\leq 4.12=M_2$ and $M_3=1.0\leq \mu_2\leq 5.2=M_4$.
The minimizer and the maximizer of (\ref{eq:2comp_cdf}) occur at coordinates of the form
$(a,b)$, where $a$ can take the values $y_i$, $M_1$ or $M_2$, and $b$ can
take the values $y_i$, $M_3$ or $M_4$. Evaluating (\ref{eq:2comp_cdf}) at these coordinates
yields the desired minimum and the maximum. At time $t$, let $\theta_{\min,t}$ and $\theta_{\max,t}$ denote the minimizer
and the maximizer, respectively.
Minimization and maximization of (\ref{eq:2comp_cdf}) with respect to $\pi$ (assuming that $0<a\leq\pi\leq b<1$
for some $a,b$ obtained using Gibbs sampling) would have led to the independent
distribution functions $F^L_i$ and $F^U_i$, but there exists a monotonicity structure in the the conditional
distribution of $\pi$ (see also \ctn{Robert04}) which can be exploited to reduce the gaps between $F^L_i$ and $F^U_i$, by keeping $\pi$
fixed in the lower and the upper bounds. Moreover, since optimization with respect to $\pi$ is no longer
needed, truncation of the parameter space of $\pi$ is not required. Details follow.

\subsection{Monotonicity structure in the simulation of $\pi$}
\label{subsec:pi_monotonicity}

It follows from (\ref{eq:full_cond_pi}) of MB
that $\pi\sim Beta(n_1+1,n-n_1+1)$. Then, at time $t$, $\pi$ can be represented as
\begin{equation}
\pi_t=\sum_{k=1}^{n_1+1}R_{\pi,t,k}\big /\sum_{k=1}^{n+2}R_{\pi,t,k},
\label{eq:pi_t}
\end{equation}
where $\{R_{\pi,t,k};k=1,\ldots,n+2\}$ is a random
sample from $Exp(1)$, that is, the exponential distribution with mean 1. Thus, $\pi_t$ is increasing with respect to $n_1$,
since the set of random numbers is fixed for all the Markov chains at time $t$.
The form of (\ref{eq:2comp_cdf}) suggests that the distribution function is increasing with $\pi$ and hence with $n_1$.
Let $n_{1t}=\#\{i:z_{it}=1\}$, $n^L_{1t}=\#\{i:z^L_{it}=1\}$ and $n^U_{1t}=\#\{i:z^U_{it}=1\}$, and
note that $n^L_{1t}\leq n_{1t}\leq n^U_{1t}$ for any $t$.
Define
\begin{eqnarray}
\pi^L_t&=&\frac{\sum_{k=1}^{n^L_{1t}+1}R_{\pi,t,k}}{\sum_{k=1}^{n+2}R_{\pi,t,k}}\label{eq:pi_lower}\\
\pi^U_t&=&\frac{\sum_{k=1}^{n^U_{1t}+1}R_{\pi,t,k}}{\sum_{k=1}^{n+2}R_{\pi,t,k}}\label{eq:pi_upper}
\end{eqnarray}
With these, the lower and upper bounds of the distribution function of $z_i$ at time $t$ are given by
\begin{eqnarray}
F^L_i(\cdot\mid\pi^L_t,Y)&=&F_i(\cdot\mid\theta_{\min,t},\pi^L_t,Y)\label{eq:pi_lower_efficient}\\
F^U_i(\cdot\mid\pi^U_t,Y)&=&F_i(\cdot\mid\theta_{\max,t},\pi^U_t,Y)\label{eq:pi_upper_efficient}
\end{eqnarray}

Combining the above developments, we propose the following algorithm for perfect simulation in 2-component
mixture models, which is a slightly modified version of Algorithm \ref{algo:cftp_known}.
For our specific example, we must set $n=3$ in the algorithm below.
\begin{algo}\label{algo:cftp_2comp}\topline
CFTP for two-component mixtures \botline \normalfont \ttfamily
\end{algo}
\begin{itemize}
\item[(i)] For $j=1\ldots$, until coalescence of $Z$, repeat steps (ii) and (iii) below.
\item[(ii)] Define $\mathcal S_j=\{-2^j+1,\ldots,-2^{j-1}\}$ for $j\geq 2$,
and let $\mathcal S_1=\{-1,0\}$. 
For each $m\in\mathcal S_j$,
generate random numbers $R_{Z,m}$, $R_{\pi,m}$ and $R_{\Theta_2,m}$;
once generated, treat them as fixed thereafter for all iterations.
\item[(iii)] For $t=-2^j+1,\ldots,-1,0$,
\begin{enumerate}
\item[(a)] Calculate $\pi^L_t$ and $\pi^U_t$ given by (\ref{eq:pi_lower}) and (\ref{eq:pi_upper}).
For $t=-2^j+1$, set $n^L_{1t}=0$ and $n^U_{1t}=n$.
\item[(b)] For $i=1,\ldots,n$,
\begin{enumerate}
\item[1.] For $\ell=1,2$, calculate $F^L_i(\ell\mid\pi^L_t,Y)$ and $F^U_i(\ell\mid\pi^U_t,Y)$,
given by (\ref{eq:pi_lower_efficient}) and (\ref{eq:pi_upper_efficient}).
In this two-component example, these can be calculated 
exactly following the details presented in Sections \ref{subsec:bounding_chains}
and \ref{subsec:pi_monotonicity}.
\item[2.] Determine $z^L_{it}=F^{U-}_i(R_{z_{i,t}}\mid\pi^L_t, Y)$ and
$z^U_{it}=F^{L-}_i(R_{z_{i,t}}\mid\pi^U_t, Y)$.
\end{enumerate}
\end{enumerate}

\item[(iv)] If $z^L_{it^*}=z^U_{it^*}$ $\forall~ i$ and for some $t^*<0$, then
run the following Gibbs sampling steps from $t=t^*$ to $t=0$:
\begin{itemize}
\item[(a)] Let $Z^*=(z^*_1,\ldots,z^*_n)'$ denote the coalesced value of $Z$ at time $t^*$.
Given $Z^*$, draw $(\pi^*,\Theta^*_2)$ from the full conditionals (\ref{eq:full_cond_pi}),
(\ref{eq:full_cond_lambda}) and (\ref{eq:full_cond_mu}) in order,
using the corresponding random numbers already generated; in fact, $\pi^*$ will be computed
using the representation (\ref{eq:pi_t}). Thus, $(Z^*,\pi^*,\Theta^*_2)$
is the 
coalesced value of the unknown quantities at $t=t^*$.
\item[(b)] Carry forward the above Gibbs sampling chain started at $t=t^*$ till $t=0$,
simulating sequentially
from (\ref{eq:full_cond_z}), (\ref{eq:full_cond_pi}), (\ref{eq:full_cond_lambda}) and
(\ref{eq:full_cond_mu}). Again, $\pi$ will be simulated using (\ref{eq:pi_t}).
Then, the output of the Gibbs sampler obtained at $t=0$,
which we denote by $(Z_0,\pi_0,\Theta_{2,0})$, is a perfect sample
from the true target posterior distribution.
\end{itemize}
\end{itemize}
\rmfamily
\botline

\subsection{Results of perfect simulation in the two-component mixture example}
\label{subsec:2comp_results}

\begin{figure}
\begin{center}
\includegraphics[width=11cm,height=9cm]{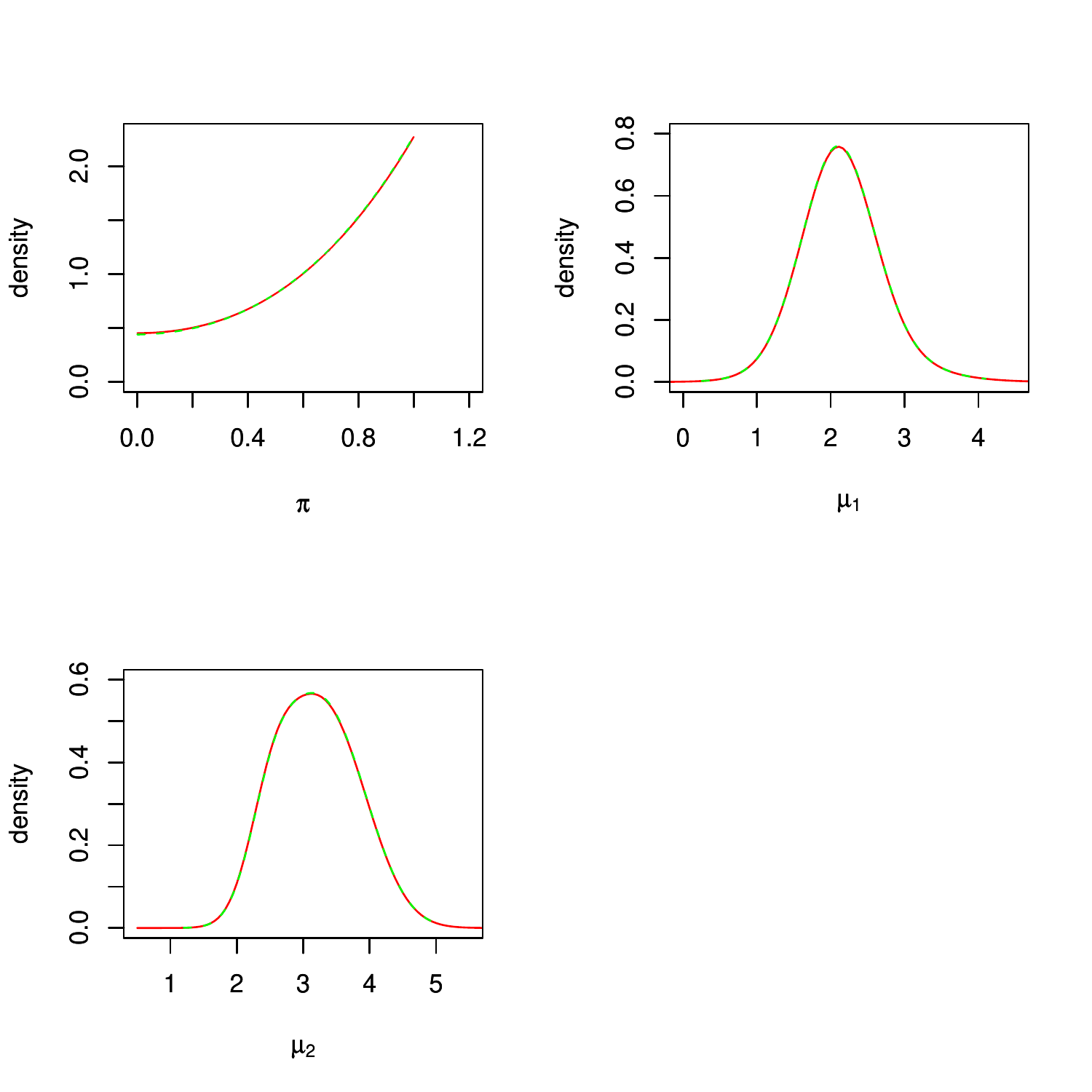}
\caption{Investigation of consequences of truncating the parameter space:
the solid and the broken lines (almost indistinguishable) correspond to the exact posterior
densities with respect to
unbounded and bounded parameter spaces, respectively. }
\label{fig:exact_com}
\end{center}
\end{figure}

We first investigated the consequences of truncating the parameter space.
Figure \ref{fig:exact_com} illustrates that in this example, the exact posterior densities of
$(\pi,\mu_1,\mu_2)$ corresponding to
bounded and full (unbounded) supports are almost indistinguishable from each other.

\begin{figure}
\begin{center}
\includegraphics[width=11cm,height=9cm]{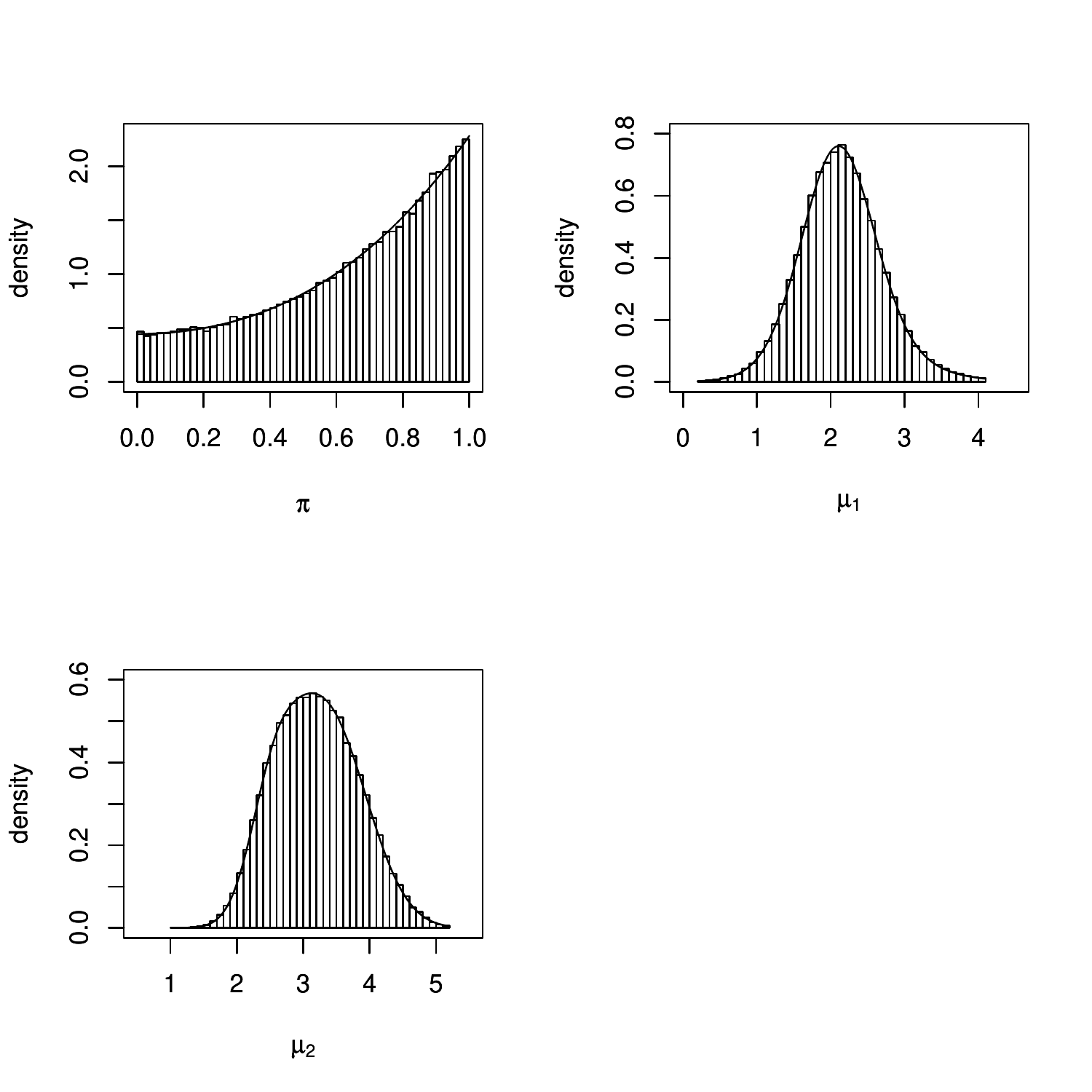}
\caption{The histograms correspond to
perfect samples drawn using our algorithm. The density lines correspond to the exact posterior density.}
\label{fig:exact_fix}
\end{center}
\end{figure}

We then implemented our perfect sampling algorithm by simulating $Z$ from the bounds (\ref{eq:pi_lower_efficient}) and (\ref{eq:pi_upper_efficient})
and simulating the upper and lower chains for $\pi$ using the formulae (\ref{eq:pi_lower}) and (\ref{eq:pi_upper}).
The histograms in Figure \ref{fig:exact_fix}, corresponding to $100,000$ $iid$ perfect samples match the exact posteriors
almost perfectly, indicating that our algorithm has worked really well.

\subsection{Comparison with perfect sampling involving simulated annealing}
\label{subsec:2comp_simulated_annealing}
In the same two-component normal mixture example, we considered two versions of our perfect sampling
algorithm: in the first version we considered exact optimization of the distribution function of $z_i$, and in the second
version we used simulated annealing for optimization. In both cases, we obtained $100,000$ $iid$ samples of $(\pi,\mu_1,\mu_2)$
at time $t=0$, using the same set of random numbers. All $100,000$ samples of the second version turned out to be equal to the
corresponding samples of the first version, suggesting great reliability of simulated annealing.

\renewcommand\baselinestretch{1.3}
\normalsize
\bibliographystyle{ECA_jasa}
\bibliography{irmcmc}

\end{document}